\documentstyle[amssymb,aps,prd,epsf]{revtex}

\begin{document}
\author{E.J. Ferrer$^{1}$, V.P. Gusynin$^{2,3}$ and V. de la Incera$^{1}$}
\address{$^{1}$ Physics Department, SUNY-Fredonia,\\
Fredonia, NY 14063, USA \\
$^{2}$Bogolyubov Institute for Theoretical Physics,\\
03143, Kiev, Ukraine, \\
$^{3}$ Department of Physics, Nagoya University,\\
Nagoya 464-8602, Japan}
\title{Thermal Conductivity in $3$D NJL Model Under External Magnetic Field}
\date{\today}
\maketitle

\begin{abstract}
The thermal conductivity of the (2+1)-dimensional NJL model in the
presence of a constant magnetic field is calculated in the
mean-field approximation and its different asymptotic regimes are
analyzed. Taking into account the dynamical generation of a
fermion mass due to the magnetic catalysis phenomenon, it is shown
that for certain relations among the theory's parameters (particle
width, temperature and magnetic field), the profile of the thermal
conductivity versus the applied field exhibits kink- and
plateau-like behaviors. We point out possible applications to
planar condensed matter. \vskip1mm PACS numbers: 11.10Kk, 11.30Qc,
11.10Wx, 12.20.Ds
\end{abstract}

\section{Introduction}

It is now well established, from the study of many relativistic
theories of massless fermions in the presence of an external
magnetic field, that a magnetic field can be a strong catalyst for
chiral symmetry breaking with the consequent generation of a
fermion dynamical mass even at the weakest attractive interaction
among fermions \cite{prl94}. This magnetic catalysis (MC) of
chiral symmetry breaking has proven to be universal, its main
features being independent of the model under consideration (see
Refs. \cite {prl94}$-$\cite{Nick} for various aspects of this
phenomenon). The universality character of the magnetic catalysis
has motivated many recent works \cite{qed4}-\cite{applic} aimed to
apply it to diverse areas of quantum physics. The essence of the
MC effect lies in the dimensional reduction of the fermion pairing
dynamics due to the confinement of these particles to their lowest
Landau level (LLL), when the pairing energy is much less than the
Landau gap $\sqrt{B}$ ($B$ is the magnitude of the magnetic field
induction). Under these circumstances, any attraction between
fermions, whenever small it might be, is strengthened by the
effective dimensional reduction in the presence of the magnetic
field, and therefore, a condensate of fermion-antifermion is
formed with the subsequent generation of a fermion mass. The
lowest LL plays in this case a role similar to that of the Fermi
surface in BCS superconductivity \cite{prl94}.

It is known that several quasiplanar systems have low-energy excitation
spectrum of quasiparticles (QP), characterized by a linear dispersion,
around the Fermi surface consisting of isolated points. The dynamics of
these QP can be described by a "relativistic" quantum field theory of
massless fermions. When such Dirac-like QP are electrically charged, they
can couple to an externally applied magnetic field which can catalyze the
condensation of QP-antiQP pairs. Then, one would expect the realization of
MC in such a kind of condensed matter systems. As a matter of fact, the MC
was suggested as the possible explanation \cite{Nick}$-$\cite{Ferrer} for
the profile of the thermal conductivity in an applied magnetic field
observed in recent experiments in planar high-$T_{c}$ cuprates \cite
{Krishana}$-$\cite{Ando2}. The MC has also been proposed \cite{K-2} as the
source of the semimetal-insulator phase transition observed in the so-called
highly oriented pyrolitic graphites (HOPG) \cite{graphite} in the presence
of a magnetic field.

Given that the heat transport is a convenient probe to understand
many basic properties of quasi-planar condensed matter systems, as
gap structure, QP density, scattering rate, etc, the study of heat
transport in a quasiplanar system subjected to MC may turn out
physically revealing. It is the goal of the present paper to study
how the MC affects the thermal transfer properties of a
(2+1)-dimensional fermion system under an applied constant
magnetic field. In particular, we calculate the thermal
conductivity of a system of QP described by a (2+1)-dimensional
Nambu-Jona-Lasinio (NJL) model which exhibits a mass (gap)
generation for fermions in the presence of a magnetic field, and
discuss possible applications to planar condensed matter systems.
To obtain the results here reported, we used the same
approximation of a constant magnetic field that was already
explored in Refs.\cite{Nick2,Semenoff}. However, our calculations
deviate considerably from what was done in these papers. Not only
we take into account the contribution of all Landau levels, but
the definition we use for the heat current itself is different. In
our formulation, when the gap induced by the magnetic field is
opened, the
thermal conductivity exhibits a new term proportional to $\sigma ^{2}$ ($%
\sigma $ is the gap). Near the phase transition point, the gap behaves like $%
\sigma \sim \sqrt{B-B_{c}}$ in the mean-field approximation. Hence, the term
proportional to $\sigma ^{2}$ yields a positive contribution in the slope of
the thermal conductivity, leading to a jump in the slope of $\kappa (B)$ at $%
B=B_{c}$ (kink-like behavior). Notice that the magnetic catalysis is mainly
responsible for the kink effect, being the critical behavior of the gap an
essential factor for this result.

We underline that to obtain such a kink-like behavior, it is
crucial to go beyond the LLL contribution in the calculation of
the thermal conductivity, since, as we show, the heat transfer
takes place due to transitions between neighboring LLs.
Physically, this is easy to understand keeping in mind that the
spatial momentum of the QP lies in the system plane. In the
presence of a magnetic field perpendicularly applied to the
two-dimensional sample, the QP spatial momentum is purely
transverse and hence quantized into discrete Landau levels.
Therefore, the transfer of kinetic energy can only occur by means
of transitions between Landau levels.

Even though our results were obtained by using a particular model, we point
out that the main outcome of the present work is of a more general and
theoretical character, as we show that the MC phenomenon can be responsible
for a kink-like effect in the thermal conductivity of a whole class of
(2+1)-dimensional relativistic fermion systems. That is, we show that the
kink effect is essentially model independent, since it is determined by the
critical behavior of the dynamically generated mass near the phase
transition point. This fact makes the basic outcome of our investigation
relevant beyond the particular model under consideration, linking it to the
universality class of theories with such a critical behavior. In connection
with this we conjecture that, since the HOPG may be described \cite{K-2} by
a model that belongs to the same universality class as that of the model
used here, the HOPG thermal conductivity in the presence of a magnetic field
should display similar kink-like behavior.

The plan of the paper is as follows. In  Sec.~\ref{thcond-general} we derive
the expression for the thermal conductivity in the (2+1)-dimensional NJL
model in the presence of a constant magnetic field and analytically study
its different asymptotics, underlying the possible application of each
result. In Sec.~\ref{numerics} we obtain the thermal conductivity vs
magnetic field profile using numerical calculations. In the reported graph,
the change of slope in the thermal conductivity profile is shown to occur at
the critical magnetic field where the fermion dynamical mass is generated at
the given temperature. The conclusions and discussion of potential
applications of our results are presented in Sec.~\ref{concl}. In Appendix~%
\ref{App-A} we derive the critical curve in the $B-T$ plane which separates
the symmetric and the symmetry-broken phases in the (2+1)-dimensional NJL
model, and study the scaling of the mass near the critical curve. A
derivation of the Kubo formula in the framework of Matsubara formalism is
given in Appendix B.

\section{Thermal conductivity in the (2+1)-D NJL model in the presence of a
constant magnetic field}

\label{thcond-general}

\subsection{\label{cond_NJL}General Results}

We start from the (2+1)-dimensional NJL Lagrangian density in an external
magnetic field
\begin{equation}
{\cal L}=\frac{1}{2}\left[ \bar{\psi}_{i},i\gamma ^{\mu }D_{\mu }\psi
_{i}\right] +\frac{g}{2N}\left( \bar{\psi}_{i}\psi _{i}\right) ^{2},
\label{lagrangian}
\end{equation}
where $D_{\mu }=\partial _{\mu }-ieA_{\mu }^{ext}$ is the covariant
derivative and the vector potential for the external magnetic field is taken
in the symmetric gauge
\begin{equation}
A_{\mu }^{ext}=\left( 0,-\frac{B}{2}x_{2},\frac{B}{2}x_{1}\right) .
\label{ext_potential}
\end{equation}
We assume that the fermions carry an additional flavor index $i=1,\dots ,N$ (%
$N=2$ for realistic $d$-wave superconductors). The Dirac $\gamma $-matrices
are taken in the reducible four-component representation.

In the absence of the bare mass term $m\bar{\psi}\psi $, the Lagrangian
density (\ref{lagrangian}) is invariant under discrete chiral symmetry
\begin{equation}
\psi \rightarrow \gamma _{5}\psi ,\quad \bar{\psi}\rightarrow -\bar{\psi}%
\gamma _{5},
\end{equation}
which forbids the fermion mass generation in perturbation theory. The
appearance of the mass (energy gap) is due to the spontaneous breaking of
the above discrete symmetry that leads to a neutral condensate of
fermion-antifermion pairs. In condensed matter physics this could correspond
to the condensation of excitons (electron-hole bound states).

Introducing the composite field $\sigma=-g(\bar\psi_i\psi_i)/N$, the
Lagrangian (\ref{lagrangian}) can be written in the form
\begin{equation}
{\cal L}=\frac{1}{2}\left[ \bar{\psi}_{i},i\gamma ^{\mu }D_{\mu }\psi
_{i}\right] -\sigma\bar\psi_i\psi_i-\frac{N\sigma^2}{2g}.
\label{lagrangian1}
\end{equation}
One readily verifies the equivalence of the Lagrangians (\ref{lagrangian})
and (\ref{lagrangian1}) by making use of the Euler-Lagrange equations (or
performing the integration over the field $\sigma$ in the functional
integral). The field $\sigma$ has no dynamics at the tree level, however, it
acquires a kinetic term due to fermion loops. The vacuum expectation value
of $\sigma$ gives a dynamical mass (gap) to fermions. The effective action
for the composite field $\sigma $ can be obtained by integrating over
fermions in the path integral (see Appendix~\ref{App-A}).

It is well known that in the absence of a magnetic field and at zero
temperature the mass generation occurs only if the coupling constant exceeds
some critical value \cite{Rosenstein}. This can be seen from the stationary
equation for the effective potential corresponding to the Lagrangian density
(\ref{lagrangian}), which in leading order of $1/N$-expansion is given by
\begin{equation}
\frac{\partial V(\sigma )}{\partial \sigma }=\sigma \left[ \sigma -\frac{%
\Lambda }{\sqrt{\pi }}+\frac{\Lambda }{\overline{g}}\right] =0
\label{stat-eq}
\end{equation}
(it follows from Eq. (\ref{gapequation}) in the limit $B=T=0$).  In (\ref
{stat-eq}) we introduced the dimensionless coupling constant $\overline{g}%
=\Lambda g/\pi $ and $\Lambda $ is the ultraviolet cutoff parameter. From
Eq. (\ref{stat-eq}) it is easy to see that there exists a critical value for
the coupling constant $\overline{g}_{c}=\sqrt{\pi }$ such that, if $%
\overline{g}<\overline{g}_{c}$ Eq. (\ref{stat-eq}) has only the trivial
solution ($\sigma =0$), while for the strong coupling limit $\overline{g}>%
\overline{g}_{c}$ a nontrivial solution ($\bar\sigma =\Lambda (\overline{g}-%
\sqrt{\pi })/$ $\overline{g}\sqrt{\pi }$) is reached which leads to the
generation of a fermion mass. Hence, the critical coupling $\overline{g}_c$
separates two phases: the weakly coupled massless phase at $\overline{g}<%
\overline{g}_c$ and the strongly coupled massive one ($\overline{g}>%
\overline{g}_c$). An applied magnetic field changes the situation
dramatically, so that the mass generation now takes place at all $\overline{g%
}>0$ \cite{prl94,njl3,njl4,Klimenko}, hence the name magnetic catalysis. At
finite $T$ and $B\neq0$ there is a critical curve in the $B-T$ plane
separating the symmetric and the symmetry broken phases (the derivation and
analysis of the critical curve in the $B-T$ plane are given in Appendix~\ref
{App-A}).

To derive an expression for the static thermal conductivity in an isotropic
system we follow the familiar linear response method and apply Kubo's
formula \cite{Ambegaokar}
\begin{equation}
\kappa =-\frac{1}{TV}{\rm Im}\int\limits_{0}^{\infty }dtt\int
d^{2}x_{1}d^{2}x_{2}\langle u_{i}(x_{1},0)u_{i}(x_{2},t)\rangle ,
\label{heat_cur_cor}
\end{equation}
where $V$ is the volume of the system, $T=1/\beta $ is the temperature, and $%
u_{i}(x,t)$ is the heat-current density operator. The brackets denote
averaging in the canonical ensemble with the density matrix $\rho =e^{-\beta
H}/Z,\,Z={\rm Tr}e^{-\beta H}$.

Physically the thermal conductivity $\kappa $ appears as a coefficient in
the equation relating the heat current to the temperature gradient
\begin{equation}
{\vec{u}}=-\kappa {\vec{\nabla}}T
\end{equation}
under the condition of absence of particle flow. If we neglect the chemical
potential the heat density coincides with the energy density, hence the
quantity that satisfies the continuity equation
\begin{equation}
\dot{\epsilon}(x)+{\vec{\nabla}\cdot }{\vec{u}(x)}=0  \label{continuity}
\end{equation}
can be interpreted as the heat current density. Equation (\ref{continuity})
defines ${\vec{u}}$ to within a divergenceless vector, which is sufficient
for calculating the conductivity. The vector ${\vec{u}}$ is obtained
automatically from the Lagrangian density (\ref{lagrangian}) as
\begin{equation}
u_{i}=\frac{\partial {\cal L}}{\partial (\partial ^{i}\psi )}\dot{\psi}+
\bar{\psi}\frac{\partial {\cal L}}{\partial (\partial ^{i}\bar{\psi})}=
\frac{i}{2}\left( \bar{\psi}\gamma _{i}\partial _{0}\psi -\partial _{0}\bar{
\psi}\gamma _{i}\psi \right) .  \label{L-derivative}
\end{equation}
When using equations of motion the last expression can be represented in the
form
\begin{equation}
u_{i}=\frac{i}{2}\left( \bar{\psi}\gamma ^{0}D_{i}\psi -\overline{D_{i}\psi }
\gamma ^{0}\psi \right).  \label{cov-derivative}
\end{equation}
At this point it is useful to underline that our definition of the heat
current does not coincide with the one used in Refs. \cite{Nick2,Semenoff}, $%
P_{i}(x)=\overline{\psi }\gamma^{0} \partial _{i}\psi -\partial _{i}
\overline{\psi }\gamma^{0} \psi $. Operator $P_{i}(x)$ cannot be obtained
from Equation (\ref{L-derivative}) by using the equations of motion, unless
the external field is zero, so it does not lead to the correct thermal
conductivity in the presence of a magnetic field. Note also, that our
quantity $u_{i}$ is explicitly gauge invariant in contrast to $P_{i}$.

The correlator of the heat currents (or polarization function) in (\ref
{heat_cur_cor}) is evaluated in the following way (see Appendix B for
details on the formalism). First, it is computed in the Matsubara finite
temperature formalism replacing the time $t$ by the imaginary time $\tau $ ($%
t=-i\tau $):
\begin{equation}
\Pi (i\Omega _{m})=\frac{1}{V}\int\limits_{0}^{\beta }d\tau e^{i\Omega
_{m}\tau }\langle T_{\tau }U_{i}(\tau )U_{i}(0)\rangle ,\quad U_{i}(\tau
)=\int d^{2}xu_{i}(x,\tau ), \quad \Omega_m=\frac{2\pi m}{\beta}.
\label{im-time_Pi}
\end{equation}
The thermal conductivity is then given by the discontinuity of the retarded
function $\Pi ^{R}(\Omega ),$ which is obtained by analytic continuation
from imaginary discrete frequencies $\Pi ^{R}(\Omega )=\Pi (i\Omega _{m}\to
\Omega +i\epsilon )$:
\begin{equation}
\kappa =\frac{1}{4Ti}\lim_{\Omega \to 0}\frac{1}{\Omega }\left[ \Pi
^{R}(\Omega +i\epsilon )-\Pi ^{A}(\Omega -i\epsilon )\right] .
\end{equation}
Neglecting vertex corrections\footnote{%
It has been argued that for small impurity densities the thermal
conductivity, unlike the electric conductivity, is unaffected by vertex
corrections \cite{Lee}.}, the calculation of the thermal conductivity
reduces to the evaluation of the bubble diagram
\begin{equation}
\Pi (i\Omega _{m})=-T\sum_{n=-\infty }^{\infty }\int \frac{d^{2}k}{(2\pi
)^{2}}tr\left[ \gamma ^{i}i\omega _{n}S(i\omega _{n},{\vec{k}})\gamma
^{i}(i\omega _{n}+i\Omega _{m})S(i\omega _{n}+i\Omega _{m},{\vec{k}})\right]
,  \label{polar_oper}
\end{equation}
where $S(\omega ,k)$ is the Fourier transform of the translation invariant
part $\tilde{S}(x-y)$ of the fermion propagator in an external magnetic
field:
\begin{equation}
S(x-y)=\exp \left( ie\int\limits_{y}^{x}A_{\lambda }^{ext}dz^{\lambda
}\right) \tilde{S}(x-y).
\end{equation}
Note, that the translation non-invariant phase of the fermion Green's
function cancels in the computation of $\Pi $. Defining a spectral
representation for $S(i\omega _{n},k)$, we can write
\begin{equation}
S(i\omega _{n},{\vec{k}})=\int\limits_{-\infty }^{\infty }\frac{d\omega
\,A(\omega ,{\vec{k}})}{i\omega _{n}-\omega }.  \label{spectr_repr}
\end{equation}
The spectral representation allows one to make analytic continuation and
find the retarded, $S^{R}$, and advanced, $S^{A}$, Green functions according
to the rule $S^{R}(\omega +i\epsilon ,{\vec{k}})=S(i\omega _{n}=\omega
+i\epsilon ,{\vec{k}})$ and $S^{A}(\omega -i\epsilon ,{\vec{k}})=S(i\omega
_{n}=\omega -i\epsilon ,{\vec{k}})$. The spectral function $A(\omega ,{\vec{%
k }})$ is given by
\begin{equation}
A(\omega ,{\vec{k}})=\frac{1}{2\pi i}\left[ S^{A}(\omega -i\epsilon ,{\vec{k}
})-S^{R}(\omega +i\epsilon ,{\vec{k}})\right] =-\frac{1}{\pi }{\rm Im}%
S^{R}(\omega +i\epsilon ,{\vec{k}}).  \label{spectrfunction}
\end{equation}
Plugging the spectral representation (\ref{spectr_repr}) into Eq.(\ref
{polar_oper}) the sum over Matsubara frequencies can be performed \footnote{%
There is a subtle point here since the sum over frequencies appears to be
divergent. However, as was shown in \cite{Ambegaokar}, this divergence
results from an improper treatment of time derivatives inside the
time-ordered product of currents in (\ref{im-time_Pi}). This divergence
disappears when the problem is treated more carefully. The prescription is
simply to ignore it.}. After this has been done, we can continue the
external frequencies to the real axis to get $\Pi ^{R}(\Omega )$. Finally,
we arrive at the following expression for the thermal conductivity
\begin{equation}
\kappa =\frac{1}{32\pi T^{2}}\int\limits_{-\infty }^{\infty }\frac{d\omega
\omega ^{2}}{\cosh ^{2}\frac{\omega }{2T}}\int d^{2}k{\rm tr}\left[ \gamma
^{i}A(\omega ,{\vec{k}})\gamma ^{i}A(\omega ,{\vec{k}})\right] .
\label{kappathroughA}
\end{equation}
In order to compute $\kappa $ we have to specify now the fermion propagator.
Since the fermion mass is generated in the (2+1)-dimensional NJL model in a
magnetic field already at weak coupling, we take the standard expression for
the massive fermion propagator in a magnetic field, decomposed over the
Landau level poles \cite{njl3,Chodos}
\begin{equation}
S(\omega ,{\vec{k}})=e^{-\frac{{\vec{k}}^{2}}{eB}}\sum\limits_{n=0}^{\infty
}(-1)^{n}\frac{D_{n}(\omega ,{\vec{k}})}{\omega ^{2}-\sigma ^{2}-2eBn},
\end{equation}
where
\begin{equation}
D_{n}(\omega ,k)=2(\omega \gamma ^{0}+\sigma )\left[ P_{-}L_{n}\left( \frac{%
2 {\vec{k}}^{2}}{eB}\right) -P_{+}L_{n-1}\left( \frac{2{\vec{k}}^{2}}{eB}
\right)\right] +4{\vec{k}}{\vec{\gamma}}L_{n-1}^{1}\left( \frac{2{\vec{k}}%
^{2}}{eB} \right)
\end{equation}
with $P_{\pm }=(1\pm i\gamma ^{1}\gamma ^{2})/2$ being projectors and $%
L_{n},L_{n}^{1}$ Laguerre's polynomials ($L_{-1}^{1}\equiv0$). Here $\sigma$
is the fermion dynamical mass obtained from the finite temperature gap
equation in a constant magnetic field (see Appendix~\ref{App-A}).

The spectral function according to (\ref{spectrfunction}) is found to be
\begin{equation}
A(\omega ,{\vec{k}})=e^{-\frac{{\vec{k}}^{2}}{eB}}\frac{\Gamma }{2\pi }
\sum\limits_{n=0}^{\infty }\frac{(-1)^{n}}{M_{n}}\left[ \frac{(\gamma
^{0}M_{n}+\sigma )f_{1}({\vec{k}})+f_{2}({\vec{k}})}{(\omega
-M_{n})^{2}+\Gamma ^{2}}+\frac{(\gamma ^{0}M_{n}-\sigma )f_{1}({\vec{k}}
)-f_{2}({\vec{k}})}{(\omega +M_{n})^{2}+\Gamma ^{2}}\right] ,
\label{spect-fun-magfield}
\end{equation}
where $M_{n}=\sqrt{\sigma ^{2}+2eBn}$ and
\begin{equation}
f_{1}({\vec{k}})=2\left[ P_{-}L_{n}\left( \frac{2{\vec{k}}^{2}}{eB}\right)
-P_{+}L_{n-1}\left( \frac{2{\vec{k}}^{2}}{eB}\right) \right] ,\quad f_{2}({\
\vec{k}})=4{\vec{k}}{\vec{\gamma}}L_{n-1}^{1}\left( \frac{2{\vec{k}}^{2}}{eB}
\right).  \label{expr:f1-f2}
\end{equation}
Here, we introduced the width $\Gamma $ of the quasiparticles, which is due
to interaction processes, in particular, scattering on impurities, having
replaced $\epsilon $ in (\ref{spectrfunction}) by a finite $\Gamma $. In
general, the scattering rate $\Gamma$, which is defined through the fermion
self-energy, $\Gamma(\omega)= -{\rm Im}\Sigma^R(\omega)$, is a
frequency-dependent quantity (as well as temperature and field dependent).
It must be determined, like the dynamical mass, self-consistently from the
Schwinger-Dyson equations. At low temperatures we are interested in its
value at $\omega=0$, so we will consider it as a phenomenological parameter.

In the absence of a magnetic field ($B=0$) it can be shown that the spectral
function (\ref{spect-fun-magfield}) reduces to
\begin{eqnarray}
A(\omega ,{\vec{k}})=\frac{\Gamma}{2\pi E}\left[ \frac{\gamma^0E-{\vec k}{%
\vec\gamma} +\sigma}{(\omega-E)^2+\Gamma^2}+ \frac{\gamma^0E+{\vec k}{\vec%
\gamma} -\sigma}{(\omega+E)^2+ \Gamma^2}\right], \quad E=\sqrt{{\vec k}%
^2+\sigma^2},
\end{eqnarray}
hence, the retarded fermion Green's function is
\begin{eqnarray}
S^R(\omega, {\vec k})=\frac{\gamma_0(\omega+i\Gamma)-{\vec k}{\vec\gamma}%
+\sigma}{(\omega+i\Gamma)^2-{\vec k}^2-\sigma^2}.
\end{eqnarray}
Straightforward calculation of the trace in (\ref{kappathroughA}), with $%
A(\omega ,{\vec{k}})$ from Eqs.(\ref{spect-fun-magfield}),(\ref{expr:f1-f2})
gives
\begin{eqnarray}
&&{\rm tr}\left[ \gamma ^{i}A(\omega ,{\vec{k}})\gamma ^{i}A(\omega ,{\vec{k}%
})\right] =  \nonumber \\
&&-\frac{16\Gamma ^{2}N}{\pi ^{2}}e^{-2{\vec{k}}^{2}/eB}\sum
\limits_{n,m=0}^{\infty }(-1)^{m+n+1}\frac{(\omega ^{2}+M_{n}^{2}+\Gamma
^{2})(\omega ^{2}+M_{m}^{2}+\Gamma ^{2})-4\omega ^{2}\sigma ^{2}}{[(\omega
^{2}+M_{n}^{2}+\Gamma ^{2})^{2}-4\omega ^{2}M_{n}^{2}][(\omega
^{2}+M_{m}^{2}+\Gamma ^{2})^{2}-4\omega ^{2}M_{m}^{2}]}  \nonumber \\
&&\times \left( L_{n}\left( \frac{2{\vec{k}}^{2}}{eB}\right) L_{m-1}\left(
\frac{2{\vec{k}}^{2}}{eB}\right) +L_{n-1}\left( \frac{2{\vec{k}}^{2}}{eB}%
\right) L_{m}\left( \frac{2{\vec{k}}^{2}}{eB}\right) \right) .
\end{eqnarray}
Performing now the integration over momenta in Eq. (\ref{kappathroughA})
produces Kronnecker's delta symbols $\delta _{n,m-1}+\delta _{m,n-1}$ due to
the orthogonality of Laguerre's polynomials, thus we get
\begin{equation}
\kappa =\frac{eB\Gamma ^{2}N}{\pi^{2}T^{2}} \sum\limits_{n=0}^{\infty
}\int\limits_{0}^{\infty }\frac{d\omega \omega ^{2}}{\cosh ^{2}\frac{\omega
}{2T}}\frac{(\omega ^{2}+M_{n}^{2}+\Gamma ^{2})(\omega
^{2}+M_{n+1}^{2}+\Gamma ^{2})-4\omega ^{2}\sigma ^{2}}{[(\omega
^{2}+M_{n}^{2}+\Gamma ^{2})^{2}-4\omega ^{2}M_{n}^{2}][(\omega
^{2}+M_{n+1}^{2}+\Gamma ^{2})^{2}-4\omega ^{2}M_{n+1}^{2}]}.  \label{kappa}
\end{equation}
Note that the factor $eB$ in front of the right hand side of (\ref{kappa})
originated from integrating over transverse momenta and gives the degeneracy
of Landau levels (more exactly, the degeneracy is $NeB/2\pi$ for the lowest
LL, $n=0$, and $NeB/\pi$ for levels with $n\geq1$). We stress the important
result that because of the appearance of the mentioned Kronnecker deltas
only transitions between neighboring Landau levels contribute into the heat
transfer. Had we restrict ourselves to the LLL as in \cite{Semenoff}, we
would have gotten zero result for $\kappa $.

Further summation over $n$ in Eq. (\ref{kappa}) can be performed expanding
the integrand in terms of partial fractions. The resulting sums are
expressed through digamma functions by means of the formula
\begin{equation}
\sum\limits_{n=0}^\infty\left[\frac{A}{n+a}+\frac{B}{n+b}+\frac{C}{n+c} +
\frac{D}{n+d}\right]=-\left[A\psi(a)+B\psi(b)+C\psi(c)+D\psi(d)\right],
\end{equation}
where for convergency $A+B+C+D=0$.

After some algebraic manipulations the final expression for $\kappa$ is
written as follows
\begin{eqnarray}
\kappa&=&\frac{N\Gamma^2}{2\pi^2T^2}\hspace{-0.5mm} \int\limits_0^\infty
\hspace{-1mm}\frac{d\omega \omega^2}{\cosh^2\frac{\omega}{2T}}\frac{1}{
(eB)^2+(2\omega\Gamma)^2} \left\{2\omega^2+\frac{ (\omega^2+\sigma^2+
\Gamma^2)(eB)^2-2\omega^2(\omega^2-\sigma^2+ \Gamma^2)eB}{%
(\omega^2-\sigma^2-\Gamma^2)^2+4\omega^2\Gamma^2}\right.  \nonumber \\
&-&\left.\frac{\omega(\omega^2-\sigma^2+\Gamma^2)}{\Gamma}{\rm Im} \psi\left(%
\frac{\sigma^2+\Gamma^2-\omega^2-2i\omega\Gamma}{2eB}\right) \right\}.
\label{kappamaggeneral}
\end{eqnarray}

This formula is the main result of our paper. Note that it is independent of
the particular model used to describe the QP interactions unless we specify
the dependence of the dynamical mass on $\Gamma,T, eB$ for a concrete model.

Another representation of Eq.(\ref{kappamaggeneral}) which is particularly
convenient for studying the small width limit $\Gamma\ll T, \sqrt{eB}$ is
\begin{eqnarray}
\kappa &=&\frac{N\Gamma ^{2}}{4\pi ^{2}T^{2}}\int\limits_{-\infty }^{\infty
} \frac{d\omega \omega ^{2}}{\cosh ^{2}\frac{\omega }{2T}}\frac{1}{(eB)^{2}
+(2\omega \Gamma )^{2}}\left\{ 2\omega^{2}+\frac{\frac{(eB)^2}{2}%
+eB\omega(\omega+\sigma)} {(\omega+\sigma)^2+\Gamma^2}+\frac{\frac{(eB)^2}{2}%
+eB\omega(\omega-\sigma)} {(\omega-\sigma)^{2}+\Gamma^2}\right.  \nonumber \\
&+&\left. eB\omega\sum\limits_{n=1}^{\infty }\frac{1}{M_n}\left[ \frac{%
\sigma^2+M_n^2+2\omega M_n}{(\omega+M_n)^2+\Gamma^2}+ \frac{2\omega
M_n-\sigma^2-M_n^2}{(\omega-M_n)^2+\Gamma^2}\right] \right\} .
\label{kappa_small_Gamma}
\end{eqnarray}
It is obtained from (\ref{kappamaggeneral}) if one takes the series
representation of the $\psi $-function in the integrand of (\ref
{kappamaggeneral}) and writes the expression in the curved brackets in
fractions of $1/(\Gamma^2+x^2)$.

We are now in the position to study different asymptotic regimes defined by
different relations among the dimensional parameters $\sigma,\Gamma,T,B$.

\subsection{Zero Magnetic Field}

First, we consider the limit of vanishing magnetic field ($B=0$). For that,
one can use the formula for the asymptotic of $\psi $-function at large
values of the argument
\begin{equation}
\psi (z)=\log z-\frac{1}{2z}-\frac{1}{12z^{2}}+\frac{1}{120z^{4}}+O\left(
\frac{1}{z^{6}}\right)
\end{equation}
to get
\begin{equation}
\kappa _{0}=\frac{N}{4\pi ^{2}T^{2}}\int\limits_{0}^{\infty }\frac{d\omega
\omega ^{2}}{\cosh ^{2}\frac{\omega }{2T}}\left[ 1+\frac{\omega
^{2}-\sigma^2+\Gamma ^{2}}{2\omega \Gamma }\left(\frac{\pi}{2}-\arctan
\frac {\sigma^2+\Gamma^2-\omega^2 }{2\omega\Gamma}\right)\right].
\label{thermcondfree}
\end{equation}

This expression describes the behavior of the thermal conductivity as a
function of temperature for massive Dirac's particles and is relevant for
the supercritical phase of the NJL model, where the mass is generated
spontaneously even at zero magnetic field.

The last expression can be evaluated analytically in two regimes, $\Gamma
\ll T$:
\begin{eqnarray}
\kappa_0\simeq\frac{N}{8\pi T^{2}\Gamma} \int\limits_{\sigma}^{\infty
}d\omega \frac{\omega\left( \omega^{2}-\sigma ^{2}\right) }{\cosh ^{2}\left(
\omega /2T\right)} \simeq\frac{N\sigma^2}{\pi\Gamma}e^{-\frac{\sigma}{T}%
},\quad T<<\sigma,  \label{thermcondmass}
\end{eqnarray}
and $\Gamma \gg T$:
\begin{eqnarray}
\frac{\kappa _{0}}{T}\simeq\frac{N}{3}\left[\frac{\Gamma^2}{\sigma^2+\Gamma^2%
} +\frac{7\pi ^{2}}{15}\frac{T^2\Gamma^2(\Gamma^2+5\sigma^2)} {%
(\sigma^2+\Gamma^2)^3}\right].  \label{thcond-T0-mass}
\end{eqnarray}

Accordingly, in the weak coupling phase of the NJL model where the dynamical
mass is not generated we obtain:
\begin{equation}
\frac{\kappa _{0}}{T}=\frac{N}{\pi }\left[ \frac{9\zeta (3)}{4}\frac{T}{%
\Gamma }+\frac{\ln 2}{2}\frac{\Gamma }{T}\right], \quad \Gamma \ll T,
\label{kappagamma<T}
\end{equation}
and
\begin{equation}
\frac{\kappa _{0}}{T}=\frac{N}{3}\left[ 1+\frac{7\pi ^{2}}{15}\frac{T^{2}}{%
\Gamma ^{2}}\right],\quad \Gamma \gg T .  \label{universal_kappa}
\end{equation}

Eqs. (\ref{kappagamma<T}), (\ref{universal_kappa}) up to an overall factor
coincide with the corresponding expressions obtained in Ref.\cite{Franz} for
the vortex state of nodal quasiparticles in the $d-$wave superconducting
phase of high-$T_c$ cuprates. The overall factor there equals $%
(v_{F}^{2}+v_{\Delta }^{2})/v_{F}v_{\Delta } $ where $v_{F},v_{\Delta }$ are
respectively the velocities perpendicular and tangential to the Fermi
surface. They originate from the quasiparticle excitation spectrum in the
vicinity of the gap nodes which takes the form of an anisotropic Dirac cone $%
E(k)=\sqrt{v_{F}^{2}k_{1}^{2}+v_{\Delta }^{2}k_{2}^{2}}$. With the overall
factor replacing $N$ ($=2$ in real $d-$wave superconductor), the first term
in Eq. (\ref{universal_kappa}) reproduces the universal (or residual)
thermal conductivity at low $T$ in the so-called ``dirty'' limit, $T\ll
\Gamma $, since it is independent of the impurity density, thus it will not
depend on the specific characteristics of the scattering processes in the
sample. The residual conductivity was recently observed in experiments \cite
{univ-kappa_exp} confirming the existence of gapless quasiparticles in $d$%
-wave cuprates at $T<T_{c}$. Note, however, that in contrast to what was
claimed in Ref.\cite{Franz2}, the low-temperature thermal conductivity for
massive quasiparticles (Eq.(\ref{thcond-T0-mass})) does not exhibit a
universal behavior when $T\to0$. This peculiarity of the low-temperature
thermal conductivity can be used to find out experimentally the second gap
in cuprates.

Expressions (\ref{kappagamma<T}), (\ref{universal_kappa}) were recently used
to propose a scenario for the arising of a plateaux at high magnetic fields
\cite{Franz} in Krishana's experiment. In that scenario the width $\Gamma $
of QPs becomes dependent on the field due to scattering on disordered
vortices, thus $\Gamma $ becomes $\Gamma _{0}+\Gamma _{B}$ where the field
induced width $\Gamma _{B}$ is calculated to be $\sim \sqrt{B}$. All the
information regarding the magnetic field is encoded now in the total width $%
\Gamma ,$ hence the magnetic field is not explicitly present. If we start
with a weak magnetic field , when $\Gamma _{B}\ll \Gamma _{0}\ll T$ , the
thermal conductivity follows first the expression (\ref{kappagamma<T}) (weak
field regime) decreasing with the field. At some point, when $\Gamma $
becomes of order $T$ a crossover takes place to the high field regime (\ref
{universal_kappa}) with plateaux. Physically, such a scenario is applicable
only if there is a small number of vortices with large distances between
them and the magnetic field is basically confined in tubes. However, for the
field range of interest, $H_{c{1}}\ll B\ll H_{c{2}}$,  where $H_{c1},H_{c2}$
are the lower and upper critical magnetic fields of the high $T_{c}$
superconductor respectively, the vortices are dense enough to overlap
strongly giving rise to an effective uniform magnetic field in the whole
plane \cite{Anderson}, so the above scenario is not perhaps the most
appropriate for this field range.

\subsection{Non-zero Magnetic Field}

We shall analyze now the thermal conductivity in the presence of a uniform
magnetic field in the whole plane. The analysis in all cases will be made at
fixed $T$ and $\Gamma $, and we do not assume the last one to be dependent
on the field.

From the phase transition analysis of the (2+1)-dimensional NJL model (see
Appendix~\ref{App-A}) it follows that at finite temperature there exists a
critical value of the magnetic field $B_{c}(T)$, above which the magnetic
catalysis phenomenon occurs generating a dynamical fermion mass even at weak
coupling (in what follows we consider only the weak coupling case $g\alt %
g_{c}$). For magnetic fields less than the critical one ($eB_{c}(T)\sim
16T^{2}$) the dynamical mass is zero ($\sigma =0$).

\subsubsection{Narrow Width}

To study the narrow width limit $\Gamma \to 0$, we replace the fractions $%
\Gamma /(\Gamma ^{2}+x^{2})$ in Eq.(\ref{kappamaggeneral}) by $\pi \delta
(x).$ Then, after integrating over $\omega $, what is equivalent to
evaluating in the mass shell for the different LL, we obtain
\begin{equation}
\kappa \simeq \frac{N\Gamma }{4\pi T^{2}}\left\{ \frac{(eB)^{2}}{%
(eB)^{2}+4\sigma ^{2}\Gamma ^{2}}\cdot \frac{\sigma ^{2}}{\cosh ^{2}\frac{
\sigma }{2T}}+\sum\limits_{n=1}^{\infty }\frac{(2eB)^{2}n(\sigma ^{2}+2eBn)}{
(eB)^{2}+4(\sigma ^{2}+2eBn)\Gamma ^{2}}\cdot \frac{1}{\cosh ^{2}\frac{\sqrt{
\sigma ^{2}+2eBn}}{2T}}\right\} ,\quad \Gamma \to 0  \label{kappagammato0}
\end{equation}
In (\ref{kappagammato0}) we kept $\Gamma ^{2}$ in the denominators in order
to be able to reproduce a smooth behavior of $\kappa (B)$ in the limit $B\to
0$. The origin of the first term in (\ref{kappagammato0}), whose essential
role in the kink-like behavior is discussed below, can be traced back to the
leading contribution of the zeroth to first LL transitions. Note that it
contributes only when the dynamical mass is present, i.e. if $\sigma \neq 0$
. That is, because the ratio $\Gamma /((\omega -\sigma )^{2}+\Gamma ^{2})$
appearing in Eq.(\ref{kappamaggeneral}) becomes $\pi \delta (\omega -\sigma
) $ in this limit, it does not contribute to (\ref{kappagammato0}) unless $%
\sigma \neq 0,$ due to the presence of the factor $\omega ^{2}$ in the
integrand of Eq.(\ref{kappamaggeneral}). This means that in the narrow width
limit the magnetic catalysis is not only connected to the generation of the
mass, but it is responsible also for the enhancement of the transitions
between zeroth and first LLs. The fact that the thermal conductivity is
proportional in this limit to the scattering rate (width) $\Gamma $ means
that it results from transitions of quasiparticles between cyclotron orbits
mediated by scattering of QPs on impurities.

\paragraph{Weak Field Limit,}

$\sqrt{eB}<4T$.

Taking into account that at weak coupling no dynamical mass is generated for
fields below the critical value (${eB}<eB_{c}$), we take $\sigma =0$ in the
calculation that follows. Making use of the Euler-MacLaurin formula
\begin{equation}
\frac{1}{2}F(0)+\sum\limits_{n=1}^{\infty }F(n)\simeq
\int\limits_{0}^{\infty }dxF(x)-\frac{1}{12}F^{\prime }(0),
\end{equation}
expression (\ref{kappagammato0}) with zero dynamical mass can be recast for
subcritical fields $\sqrt{ eB}<\sqrt{eB_{c}}\sim 4T$ in the form
\begin{equation}
\kappa =\frac{2NT^{2}}{\pi \Gamma }\int\limits_{0}^{\infty }\frac{dxx^{5}}{
\cosh ^{2}x}\frac{1}{x^{2}+\left( \frac{eB}{4T\Gamma }\right) ^{2}}.
\label{kappa_sub_crit}
\end{equation}
Eq.(\ref{kappa_sub_crit}) shows monotonic decreasing of $\kappa $ with
increasing magnetic field $B$ (at $B=0$ it reproduces the leading term in
Eq.(\ref{kappagamma<T})). Note that the scale $\sqrt{4T\Gamma }$ marks the
crossover point where the transition from superweak ($\sqrt{eB}\alt\sqrt{%
4T\Gamma}$) to weak ($\sqrt{4T\Gamma }\alt\sqrt{eB}<4T$) fields takes place.

\paragraph{Strong Field Limit,}

$\sqrt{eB}\agt 4T$.

We shall consider now the strong field regime, $\sqrt{eB}\agt 4T$, where a
nonzero dynamical fermion mass is generated in the weakly interacting system
(we are interested mainly in the region of coupling constants $\overline{g}%
\alt \overline{g}_{c}$ where the scaling $\sigma\sim\sqrt{eB}$ is achieved).

Let us start, however, analyzing the case of free massless fermions ($%
\sigma=0$). In this case, one can use Eq. (\ref{kappagammato0}), after
evaluating it in $\sigma =0$, to describe the very large field ($\sqrt{eB}%
\gg 4T$) behavior of $\kappa$ in the narrow width case, what yields an
exponential fall of the conductivity
\begin{equation}
\kappa \simeq \frac{8N\Gamma eB}{\pi T^{2}}e^{-\frac{\sqrt{2eB}}{T}}.
\label{kappa_free_large_B}
\end{equation}
Coming back to the interacting case and after dropping the term depending on
$\Gamma$ in the denominators of Eq.(\ref{kappagammato0}), we obtain
\begin{equation}
\kappa \simeq \frac{N\Gamma }{4\pi T^{2}}\left\{ \frac{\sigma^{2}} {\cosh^{2}%
\frac{\sigma }{2T}}+\sum\limits_{n=1}^{\infty}\frac{4n(\sigma^{2}+2eBn)} {%
\cosh ^{2}\frac{\sqrt{\sigma ^{2}+2eBn}}{2T}}\right\}.  \label{kappa_large_B}
\end{equation}
In the limit of large fields ($\sqrt{eB}\gg 4T),$ the first term in (\ref
{kappa_large_B}) is the leading one. Such a term would produce a sharp
plateau, were the fermion mass a constant. Since in the model under
consideration the mass is dynamical and it behaves as $\sigma \sim \sqrt{eB}$
at $\sqrt{eB}\gg \sqrt{eB_{c}}\simeq 4T,$ the first term gives rise to an
exponential decrease as in the case of free massless fermions (\ref
{kappa_free_large_B}) for asymptotically large fields.

\paragraph{Near the phase transition point,}

$eB\agt eB_{c}$.

In this case, the thermal conductivity can still be approximated by Eq.(\ref
{kappa_large_B}). As shown in Appendix A, near the mean field phase
transition point $\sigma \approx \frac{1}{2}\sqrt{eB-eB_{c}},$ so if the field lies in the
interval $eB_{c}<eB\alt 2eB_{c}$, the dynamical mass $\sigma \lesssim 2T$
and hence the $\cosh ^{2}\frac{\sigma }{2T}$ appearing in the first term of
Eq.(\ref{kappa_large_B}) is of order one. In this field region the first
term of Eq.(\ref{kappa_large_B}) gives a positive contribution to the
derivative of $\kappa $ close to the critical point. That positive
contribution leads to a jump in the slope of $\kappa $ at $eB=eB_{c}$
therefore showing a kink-like behavior for $\kappa $ in the narrow-width
case.

\subsubsection{Finite Width}

Let us consider now the case where the width $\Gamma $ is small but finite.
We are particularly interested in the behavior of $\kappa $ near the phase
transition point, where $eB\gtrsim eB_{c},$ and therefore $\sqrt{eB}>4T$, $%
\Gamma ,\sigma .$ From Eq.(\ref{kappa}) one can see that for these fields
the contribution of transitions between Landau levels with $n\ge 1$ in the
integrand behaves as $1/(eB)^{2},$while the transitions between zeroth and
first LL decrease as $\sim 1/eB$. However, the LL degeneracy is also
proportional to $eB,$ what implies that the transitions between the zeroth
and first LL are not suppressed despite the fact that the gap between levels
grows with the field. The leading in $1/eB$ behavior is easy to obtain from
Eq.(\ref{kappa_small_Gamma}). It is given by the expression
\begin{equation}
\kappa =\frac{N\Gamma ^{2}}{4\pi ^{2}T^{2}}\int\limits_{0}^{\infty }\hspace{
-1mm}\frac{d\omega \omega ^{2}}{\cosh ^{2}\frac{\omega }{2T}}\left\{ \left(
1+\frac{2\omega (\omega +\sigma )}{eB}\right) \frac{1}{(\omega +\sigma
)^{2}+\Gamma ^{2}}+\left( 1+\frac{2\omega (\omega -\sigma )}{eB}\right)
\frac{1}{(\omega -\sigma )^{2}+\Gamma ^{2}}\right\}
\label{Leading_B_behavior}
\end{equation}
As discussed in the previous subsection, near the transition the dynamical
mass behaves as $\sigma \approx \frac{1}{2}\sqrt{eB-eB_{c}},$ so we can
expand Eq.(\ref{Leading_B_behavior}) around $\sigma =0$ to obtain
\begin{equation}
\kappa =\frac{N\Gamma ^{2}}{2\pi ^{2}T^{2}}\int\limits_{0}^{\infty }\hspace{
-1mm}\frac{d\omega \omega ^{2}}{\cosh ^{2}\frac{\omega }{2T}}\left\{ \frac{%
1+2\omega ^{2}/eB}{\omega ^{2}+\Gamma ^{2}}+\frac{\sigma^{2}}{
(\omega^{2}+\Gamma ^{2})^{2}}\left[\frac{3\omega^{2}-\Gamma^2} {%
\omega^{2}+\Gamma ^{2}}+\frac{2\omega ^{2}}{eB} \left(\frac{%
\omega^{2}-3\Gamma^2}{\omega^{2}+\Gamma^{2}}\right) \right] \right\}.
\label{kappa_sigma expansion}
\end{equation}
Clearly, for $\Gamma <\sqrt{3}(2T),$ the term proportional to $\sigma ^{2}$
gives a positive contribution to the derivative of $\kappa $ with respect to$%
\ B$ near the transition point, so there is a jump in the slope of $\kappa $
at the critical point: a kink-like effect. Notice that if $\sigma $ were
zero or constant, the derivative of the thermal conductivity would satisfy $%
\frac{d\kappa }{deB}\simeq \frac{-C}{\left( eB\right) ^{2}},$with C
positive, so no kink-like effect would be present. On the other hand, since $%
\sigma $ is dynamical, near the transition point $\frac{d\kappa }{deB}\simeq
$ $\alpha -\beta /eB+O((eB)^{-2}),$ with $\alpha $ and $\beta $ positive,
hence one can see that the dynamical mass not only allows for a jump in the
slope, but it flattens the profile after the critical field, allowing at
least for certain region of the parameter space a behavior of almost zero
slope: a plateau-like profile. This kink and plateau -like behavior is
corroborated by numerical calculations in the next Section.

We highlight that the dynamical mass is needed to obtain the kink-like
effect in both narrow and finite width cases. Hence, in our model the kink
of the thermal conductivity is directly linked to the magnetic catalysis
phenomenon. Moreover, any model with the same critical behavior for $\sigma $
would lead to a similar effect. This means that our results are indeed model
independent, since any relativistic theory of interacting fermions that
belongs to the universality class determined by the critical behavior here
considered would yield a similar kink-like feature.

One should note that the mean field behavior of the dynamical mass
$\sigma\sim\sqrt{eB-eB_c}$ may change if higher order corrections
(fluctuations) are taken into account in the gap equation. The
fluctuations could either change the phase transition to a first order
one, with a discontinuity in $\sigma$ at the phase transition point
(this was a suggestion made by Laughlin in \cite{Laughlin}), or to a
non-mean-field order phase transition, with the scaling law
$\sigma\sim(eB-eB_c)^\nu$ where $\nu> 1/2$. While in the former case a
discontinuity will appear in the thermal conductivity, in the latter
case the conductivity will be a smooth function of the magnetic field,
and a singularity will move to its higher derivatives.

\subsubsection{Low temperature limit}

Finally, we give an expression for the thermal conductivity when the
temperature is much less than both $\Gamma$ and $eB$. At low $T$ the
function $cosh^{-2}(\omega/2T)$ in Eq.(\ref{kappamaggeneral}) is very
sharply peaked at $\omega=0$, thus, expanding the rest of the integrand over
$\omega$ and performing the integration we get
\begin{eqnarray}
\kappa&=&\frac{NT}{3}\Big\{\frac{\Gamma^2}{\sigma^2+\Gamma^2}+\frac{%
7\pi^2T^2 \Gamma^2}{5}\left[\frac{3\sigma^2-\Gamma^2}{(\sigma^2+\Gamma^2)^3}
+\frac{2}{eB}\frac{\sigma^2-\Gamma^2}{(\sigma^2+\Gamma^2)^2}\right.
\nonumber \\
&+&\left.\frac{2}{(eB)^2} \frac{\sigma^2-\Gamma^2}{\sigma^2+\Gamma^2}-\frac{%
\sigma^2-\Gamma^2}{(eB)^3} \psi^\prime\left(\frac{\sigma^2+\Gamma^2}{2eB}%
\right)\right]\Big\}.  \label{kappa-low-T}
\end{eqnarray}
It is easy to see, using the asymptotic of the $\psi-$function at large
values of its argument, that when $B\to0$ the expression (\ref{kappa-low-T})
goes to Eq.(\ref{thcond-T0-mass}) in spite of the fact that we  approach $B=0
$ from the side $B>T$. For large fields we get
\begin{eqnarray}
\kappa&=&\frac{NT}{3}\Big\{\frac{\Gamma^2}{\sigma^2+\Gamma^2}+\frac{%
7\pi^2T^2 \Gamma^2}{5}\left[\frac{3\sigma^2-\Gamma^2}{(\sigma^2+\Gamma^2)^3}
-\frac{2}{eB}\frac{\sigma^2-\Gamma^2}{(\sigma^2+\Gamma^2)^2}\right]\Big\}.
\label{small-T-large-B}
\end{eqnarray}
Note that at large $eB$ the thermal conductivity would approach some
constant value in the case of constant mass, if $\sigma>\Gamma$, $\kappa(B)$
approaches that asymptotical value from below. This resembles the low $T$
behavior of the thermal conductivity in $d-$wave cuprates \cite{Vekhter}. On
the other hand, if the mass $\sigma$ is a dynamical one with asymptotic
behavior $\sigma\sim\sqrt{eB}$ as in our four-fermion model, then $\kappa(B)$
goes to zero as $1/eB$ at large fields. This is different from the thermal
conductivity behavior for massless particles which tends to the universal
constant $\kappa=NT/3$ for $\sigma=0$ at large $B$ (see, Eq.(\ref
{small-T-large-B})).

\section{Thermal Conductivity Profile: Numerical Calculations}

\label{numerics}

In this section we do a numerical study of the profile of the thermal
conductivity versus the applied magnetic field, taking into account the
generation of the dynamical mass at a critical field that depends on the
temperature. The field-dependence of the finite-temperature dynamical mass
is obtained from the solution of the gap equation (\ref{gapequation})
derived in Appendix~\ref{App-A}.

To numerically investigate the behavior of the thermal conductivity (\ref
{kappamaggeneral}) within a parameter range that can be of interest for
condensed matter applications, we need to restore all the model parameters,
like $\hbar ,c,k_{B},v_{F},v_{\Delta }$. Following Ref.\cite{Liu} we write
the Lagrangian density as
\begin{equation}
{\cal L}=\frac{1}{2}[\bar{\psi}_{i},\left( i\gamma ^{0}\hbar \frac{ \partial
}{\partial t}+v\gamma ^{j}(i\hbar \frac{\partial }{\partial x^{j}}- \frac{e}{%
c}A_{j})\right) \psi _{i}]+\frac{gv}{2N}(\bar{\psi}_{i}\psi _{i})^{2},
\label{N1}
\end{equation}
where $v_{F}$ and $v_{\Delta }$ entering in $v=\sqrt{v_{F}v_{\Delta }}$ were
defined in the previous section, and the external potential is given by Eq.(%
\ref{ext_potential}). As known, this Lagrangian is equivalent to
\begin{equation}
{\cal L}=\frac{1}{2}[\bar{\psi}_{i},\left( i\gamma ^{0}\hbar \frac{ \partial
}{\partial t}+v\gamma ^{j}(i\hbar \frac{\partial }{\partial x^{i}}- \frac{e}{%
c}A_{j})\right) \psi _{i}]-\sigma v\bar{\psi}_{i}\psi _{i}-\frac{ N\sigma
^{2}v}{2g},  \label{N2}
\end{equation}
since the Euler-Lagrange equation for the auxiliary scalar field $\sigma $
obeys the constraint $\sigma =-(g/N)\bar{\psi}_{i}\psi _{i}$ so that the
Lagrangian density (\ref{N2}) reproduces Eq. (\ref{N1}) upon application of
this constraint. The effective action for the composite field $\sigma $ can
be obtained by integrating over fermions in the path integral. From the
minimum condition of the effective potential $V(\sigma )$ one finds that, at
fixed $T$, there is a critical value of the magnetic field $\sqrt{eB_{c}}%
/T\simeq 4.1476$ such that for subcritical fields $eB\leq eB_{c}$ the gap is
zero, while for $eB>eB_{c}$ it is given by the non-trivial solution of the
gap equation (\ref{gapequation}) (see Appendix~\ref{App-A}).

Notice that $\sigma v$ has dimension of energy and plays the role of $mc^{2}$
in the Dirac Lagrangian density. To generate a plot of $\kappa /\kappa _{0}$
($\kappa _{0}$ is the thermal conductivity at zero field) versus the
magnetic field, we need to substitute $T\to k_{B}T,\Gamma \to \hbar \Gamma
,eB\to (\hbar v^{2}/c)eB$, where $B$ is measured in Gauss. It is convenient
to measure all energetic quantities in degrees of ${}K$, what leads to the
replacement $eB\to 0.8\cdot 10^{10}(v/c)^{2}\,{}^{\circ }K^{2}\cdot B(Tesla)$%
, where the magnetic field is measured now in Tesla's. Thus, using the
approximated value of the characteristic velocity $v_{D}\simeq 10^{7}cm/s$
\cite{Liu},\cite{K-2}, $eB\to 2.92\cdot10^{2}\cdot {}^{\circ }K^{2}\cdot
B(Tesla)$, and we obtain the critical curve $B=0.014\cdot T^{2},$.

\begin{figure}
\epsfxsize=9cm \epsffile{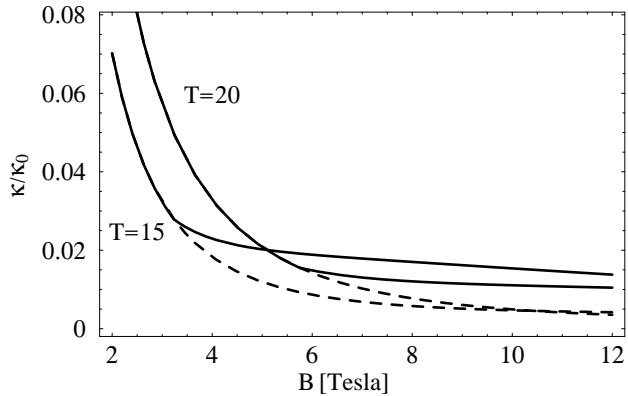}
 \caption{The magnetic
field dependence of $\kappa $ at $T=20 K$ and $T=15 K$ in the
narrow width case $(\Gamma=5 K).$ The solid lines represent
$\kappa
/\kappa _{0}$ when a QP gap $\sigma $ is MC-induced at $B\geq B_{c}(T)$ ($%
B_{c}(20)=5.75{\rm T},\,B_{c}(15)=3.23{\rm T}$). The dashed lines
represent the behavior of $\kappa /\kappa _{0}$ when $\sigma $
remains zero at $B\geq B_{c}(T)$. } \label{fig:1}
\end{figure}

Let us numerically find the profile of $\kappa $ versus the magnetic field
at a fixed $T$. In Fig.\ref{fig:1} $\kappa \left( B\right) $ has been
plotted for two different temperatures taking into account the generation of
the dynamical mass for $eB>eB_{c}$. Here we corroborate what we had already
argued in Section 2 based on the analytic result found for $\kappa $: due to
the appearance of $\sigma $ at a critical field that depends on the
temperature, $\kappa $ exhibits a kink behavior in its profile with the
magnetic field. Moreover, for $B>B_{c}$ the kink is followed by a region
where $\kappa $ is only weakly dependent on the field (plateau-like region).
With decreasing temperature, the position of the kink moves to the left in
accordance with the critical line $B_{c}=0.014T^{2}$.

The numerical calculations revealed the sensitivity of the kink-plateau
feature of the thermal conductivity to the relation between $\Gamma$ and $T$%
. Only when $\Gamma$ was not much smaller than $T$ the thermal conductivity
showed a kink-plateau profile (in Fig.\ref{fig:1} the curves shown
correspond to the ratios $\Gamma/T =0.25$ and $0.33$)).

\section{Concluding Remarks}

\label{concl} In the present paper we study the thermal
conductivity of relativistic fermions in a (2+1)-dimensional
four-fermion interaction model as a function of the applied
magnetic field, the temperature and the particle width. We have
shown that, for certain relations among these parameters, the
profile of the thermal conductivity versus the applied field
exhibits a kink-like behavior at $B\simeq B_{c}$, where $B_{c}$ is
the critical field for the generation of a fermion dynamical mass
$\sigma $, followed by a plateau-like region at $B\geq B_{c}.$ We
point out that the kink effect is the consequence of two main
features: the generation of a fermion gap in the presence of the
magnetic field (MC phenomenon), and the enhancement of the
zeroth-to-first LLs transitions.

A main outcome of our investigation is that the relevant
properties of the thermal conductivity of the (2+1)-dimensional
relativistic QP system around the critical point are model
independent. Indeed, the essential ingredient of the effective
model required to produce the kink-like effect in the thermal
conductivity is the critical behavior of the dynamical mass
induced by MC near the phase transition, so not much depends on
the concrete form of the effective Lagrangian. This fact makes our
result relevant beyond the particular model under consideration,
linking it to the universality class of theories with such a
critical behavior. Such an universal character opens a window for
possible applications.

From a quantum field theory viewpoint, condensed matter systems whose Fermi
surface are only characterized by nodal points are especially interesting
for us, since at low energies they can be described by relativistic quantum
field theory models of massless fermions \cite{Volovik}. Along that
direction, a feasible possibility for the application of our results is the
heat transport properties of graphite in the presence of a magnetic field.
Let us recall that HOPG materials \cite{graphite} have layered structure
with two isolated points in the Brillouin zone where the dispersion is
linear. Their electronic states can be thus described in terms of
relativistic charged particles \cite{Gonzales}. This graphite could exhibit
the phenomenon of MC as suggested in Ref.\cite{K-2} to explain the
semimetal-insulator phase transition observed in HOPG in the presence of a
magnetic field perpendicular to the layers. As the quasiparticles in the
graphite are subjected to Landau quantization under a perpendicularly
applied magnetic field, our results should have full strength there and we
anticipate that the thermal conductivity of these systems will show a
behavior similar to the one reported in the present work (for computation of
the electric conductivity in graphite along the lines followed in the
present work, see recent paper \cite{GGMS}).

On the other hand, the characteristic feature of $d$-wave
superconductors is also the existence of nodal points (four in
this case) where the order parameter vanishes, thus the Fermi
surface consists of four isolated points with excitations around
them being well-defined gapless quasiparticles (QP). The kinetic
part of the QPs effective Lagrangian is nothing but the Dirac
Lagrangian for two species of massless four-component spinors
\cite {condmatdirac}. At low temperatures such QPs give the main
contribution to thermodynamic and transport properties. There is
now considerable experimental evidence for the existence of
well-defined QPs in the superconducting state of cuprates (see
Ref.\cite{Randeria} and references therein).

As was mentioned in the Introduction, the MC has been suggested \cite{Nick}$-
$\cite{Liu} to be behind the odd behavior of the thermal conductivity of
high-Tc superconducting cuprates in a magnetic field observed in the
experiment \cite{Krishana}, \cite{Aubin},\cite{Ando2}, although alternative
solutions, not based on MC, has also been proposed \cite{Franz},\cite
{Laughlin}. According to the experiment done by Krishana et al. \cite
{Krishana}, and later reproduced by other authors \cite{Aubin},\cite{Ando2},
at temperatures significantly lower than $T_{c}$ of superconductivity, the
thermal conductivity $\kappa(B)$, as a function of a magnetic field
perpendicularly applied to the cuprate planes of the samples, displays a
sharp break in its slope (kink-like behavior) at a transition field $%
B_{\kappa }$, followed by a plateau region in which it ceases to change with
increasing field up to the highest attainable fields $\sim 14T$. The
critical temperature for the appearance of the kink-like behavior scales
with the magnetic field as $T_{\kappa }\sim \sqrt{B}$ similar to the scaling
of the critical temperature with the field found in NJL models \cite{njl3}.
It is worthy to mention here that the reliability of the thermal
conductivity experiments of Krishana et al. \cite{Krishana} has been a
matter of debate in the last years \cite{Ando}, although it seems that
finally the contradictory results have been clarified and understood \cite
{Ando2}.

We emphasize that none of the previously mentioned \cite{Nick}$-$\cite{Liu},%
\cite{Franz},\cite{Laughlin} attempted explanations of Krishana's
experiment were able to obtain on a theoretical basis the
kink-like behavior observed in the experiment \cite{Krishana}. In
the case of our results, although it is striking that, as
discussed in our previous paper \cite{Ferrer} and shown in
Fig.\ref{fig:1}, our numerical curves exhibit a profile of the
thermal conductivity with the applied magnetic field with similar
qualitative characteristics\footnote{%
In addition to the kink-plateau behavior and the scaling of the
critical field with the temperature, our curves have another
similarity with the experimental behavior reported in
\cite{Krishana}, namely, with decreasing $T $ the crossing of the
curves occurs in such a way that the lower $T$ curve reaches the
higher value at large fields.} to those observed in the experiment
\cite{Krishana}, we cannot claim that our results are valid in the
vortex state of the superconductor since our calculations are
based on the Landau level quantization, whose applicability to the
d-wave superconductor in the presence of vortices has been
recently subjected to intense criticism
\cite{Melnikov,Tesa,F-T-approach}.

Nevertheless, the mechanism of generating a kink-like effect in
the thermal conductivity via the MC phenomenon, as studied in the
previous sections, should be relevant for condensed matter systems
with Dirac-like charged QP on which Landau level quantization is
feasible. In this direction, we would like to point out that there
is still some chance for the realization of MC in cuprate systems.
As it has been recently argued in Refs.
\cite{Chakra},\cite{Nayak}, the physical picture underlying the
description of superconducting cuprates may involve the interplay
of two different phases with their corresponding order parameters:
one $d_{x^{2}-y^{2}}$ superconducting (DSC) and one of density
wave order (DDW). The DDW state, unlike the DSC state, does not
break gauge invariance, so the quasiparticles of the DDW state can
form Landau levels under an applied magnetic field \cite {Nayak}.
Although the above scenario is still speculative and more
experimental confirmation is required, one can venture that given
that the charged QP excitations around the nodes of the DDW order
parameter have Landau levels, they can be subjected to a MC phase
transition at some critical magnetic field, and hence our
theoretical results could be relevant in that case for the
description of the heat transport properties of the DDW-state in
the magnetic field.

\begin{acknowledgements}
{We would like to acknowledge V.A.~Miransky for useful and
stimulating discussions and to Z. Tesanovic for calling our
attention to Ref.\cite{Nayak}. We are grateful to A.~Hams and
I.~Shovkovy for help in the numerical calculation. This research
has been supported in part by the National Science Foundation
under Grant No. PHY-0070986. The work of V.P.G. is supported also
by the SCOPES-projects 7~IP~062607 and 7UKPJ062150.00/1 of the
Swiss National Science Foundation and Grant-in-Aid of Japan
Society for the Promotion of Science (JSPS) No. 11695030. He
wishes to acknowledge the JSPS for financial support. }
\end{acknowledgements}

\appendix

\section{Dynamical Mass Scaling near the Phase Transition Point}

\label{App-A}

The effective potential of the (2+1)-dimensional NJL model in a constant
external magnetic field at finite temperature was computed in \cite{njl3}.
The integration over fermion fields in the functional integral with the
Lagrangian (\ref{lagrangian1}) can be performed to give the effective
potential in the form
\begin{eqnarray}
V(\sigma )=\frac{N\sigma ^{2}}{2g}+\frac{NeB}{4\pi ^{3/2}}\int_{0}^{\infty }
\frac{dt}{t^{3/2}}e^{-t\sigma ^{2}}\coth {eBt}\Theta _{4}\Big(0|\frac{i}{
4\pi T^{2}t}\Big) =V_{0,B}(\sigma )+V_{T,B}(\sigma ),  \label{eq:VBT}
\end{eqnarray}
where the temperature independent part of the potential is given by
\begin{eqnarray}
V_{0,B}(\sigma ) &=&\frac{N\sigma ^{2}}{2g}+\frac{NeB}{4\pi ^{3/2}}%
\int_{0}^{\infty }\frac{dt}{t^{3/2}}e^{-t\sigma ^{2}}\coth {eBt}  \nonumber
\\
&=&\frac{N}{\pi }\Bigg[\frac{1}{2}M_{0}\sigma ^{2}-\sqrt{2}(eB)^{3/2}\zeta
\left( -\frac{1}{2},\frac{\sigma ^{2}}{2eB}+1\right) -\frac{\sigma eB}{2} %
\Bigg],  \label{eq:poten41}
\end{eqnarray}
(the mass scale parameter $M_{0}=\pi /g-\Lambda /\sqrt{\pi }$ and $\Lambda $
is the ultraviolet cutoff which is taken much bigger than all other
parameters in the model) and the part depending on temperature is
\begin{eqnarray}
V_{T,B}(\sigma )=\frac{NeB}{4\pi ^{3/2}}\int_{0}^{\infty }\frac{dt}{t^{3/2}}%
e^{-t\sigma ^{2}}\coth {eBt}\left[ \Theta _{4}\Big(0|\frac{i}{4\pi T^{2}t}%
\Big)-1\right] .
\end{eqnarray}
Here $\theta _{4}(v|\tau )$ is the Jacobi's elliptic function.

The gap equation $dV(\sigma)/d\sigma =0$ follows from (\ref{eq:VBT})
\begin{equation}
\sigma\hspace{-1mm}\left[-\frac{M_0}{\sqrt{eB}}+\frac{1}{\sqrt{2}}%
\zeta\left( \frac{1}{2}, \frac{\sigma^2}{2eB}+1\right)+\frac{\sqrt{eB}}{%
2\sigma} \tanh \frac{\sigma}{2T}+\int\limits_0^\infty\hspace{-1mm}\frac{dt}{%
\sqrt{\pi t}} \frac{e^{-\frac{\sigma^2t}{eB}}}{e^{2t}-1}\left[\theta_4%
\left(0|\frac{ieB} { 4\pi T^2t}\right)-1\right]\hspace{-1mm}\right]\hspace{%
-1mm}=0.  \label{gapequation}
\end{equation}
The solution of this equation defines the fermion dynamical mass. At $T=0,
B=0$ the gap equation admits a nontrivial solution only if the coupling $g$
is supercritical, $g>g_c=\pi^{3/2}/\Lambda$ ($M_0<0$). The gap equation at $%
T=0, B\neq0$ was also studied in the literature (see, for example, Refs.\cite
{prl94,njl3}). It was shown that it always has a nontrivial solution at all $%
g>0$ no matter how small the magnetic field $B$ might be. In the weak
coupling phase the field-induced dynamical mass at zero temperature behaves
as $\sigma_0=eB/2M_0$ in weak fields ($\sqrt{eB}\ll M_0$), while at high
fields ($\sqrt{eB}\gg M_0$) it is $\sigma_0\simeq0.446 \sqrt{eB}$. At finite
temperature the critical line in the $B-T $ plane separating the massless
and massive phases was calculated numerically in \cite{njl3} (see also \cite
{Liu}). We shall derive here an analytical solution of the gap equation near
such a critical line.

We start, actually, with the derivation of the Landau-Ginzburg-like
potential by expanding $V(\sigma )$ in powers of $\sigma $, since near the
phase transition point $\sigma $ is small, and write
\[
V(\sigma )=V(0)-\frac{1}{2}M(T,eB)\sigma ^{2}+\frac{1}{4}\lambda
(T,eB)\sigma ^{4}.
\]
As it is accustomed for second-order phase transitions, the equality of the
coefficient $M(T,eB)$ to zero defines the phase transition curve. The region
of parameters $T,B$ where $M(T,eB)>0$ corresponds to the spontaneously
broken phase with fermions acquiring the mass, whereas the region $M(T,eB)<0$
corresponds to the massless phase. Our goal is to obtain the coefficients $%
M(T,eB),$ and $\lambda (T,eB)$. When it is done the solution of the gap
equation is given by
\begin{equation}
\sigma ^{2}=\frac{M(T,eB)}{\lambda (T_{c},eB)}.  \label{Sigmma2}
\end{equation}
From (\ref{Sigmma2}) we obtain the behavior of the dynamical mass near the
phase transition point ($M(T_{c},eB_{c})=0$). In particular, at fixed
temperature and $B\to eB_{c}$ we find
\begin{eqnarray}
\sigma ^{2}(T=T_{c},eB) &=&\frac{M(T_{c},eB)}{\lambda (T_{c},eB)}\simeq
\frac{d}{deB}\left( \frac{M(T_{c},eB)}{\lambda (T_{c},eB)}\right)
_{B=B_{c}}\left( eB-eB_{c}\right)   \nonumber \\
&\simeq &\frac{1}{\lambda (T_{c},eB_{c})}\frac{d}{deB}{M(T_{c},eB)}%
|_{eB=eB_{c}}\left( eB-eB_{c}\right) .  \label{near_crit_scaling}
\end{eqnarray}
The potential $V_{0,B}(\sigma )$ is easily expanded in powers of $\sigma ^{2}
$ . Let us turn to $V_{T,B}(\sigma )$ which we write as the sum of two terms
\begin{eqnarray}
&&V_{T,B}(\sigma )=\frac{NeB}{4\pi ^{3/2}}\int_{0}^{\infty }\frac{dt}{t^{3/2}%
}e^{-t\sigma ^{2}}\left[ \Theta _{4}\Big(0|\frac{i}{4\pi T^{2}t}\Big)%
-1\right]   \nonumber \\
&&+\frac{NeB}{4\pi ^{3/2}}\int_{0}^{\infty }\frac{dt}{t^{3/2}}e^{-t\sigma
^{2}}\left( \coth {eBt}-1\right) \left[ \Theta _{4}\Big(0|\frac{i}{4\pi
T^{2}t}\Big)-1\right] .
\end{eqnarray}
The second term in the last expression can be expanded in a series in $%
\sigma ^{2}$ since the two brackets (with cotangent and $\theta -$function)
regularize the behavior of the integrand at infinity and zero, respectively.
Thus we need to calculate the first term
\begin{eqnarray}
&&\int_{0}^{\infty }\frac{dt}{t^{3/2}}e^{-t\sigma ^{2}}\left[ \Theta _{4}%
\Big(0|\frac{i}{4\pi T^{2}t}\Big)-1\right] =2\sum\limits_{n=1}^{\infty
}(-1)^{n}\int_{0}^{\infty }\frac{dt}{t^{3/2}}e^{-t\sigma ^{2}-\frac{n^{2}}{%
4T^{2}t}}  \nonumber \\
&=&4\sqrt{2T\sigma }\sum\limits_{n=1}^{\infty }\frac{(-1)^{n}}{\sqrt{n}}%
K_{1/2}\left( \frac{n\sigma }{T}\right) =-4\sqrt{\pi }T\log \left( 1+e^{-%
\frac{\sigma }{T}}\right) ,
\end{eqnarray}
where $K_{\nu }(z)$ is a modified Bessel function ($K_{1/2}(z)=(\pi
/2z)^{1/2}e^{-z}$).

Finally, we obtain the following expressions for the coefficients
\begin{eqnarray}
M(T,B) &=&\frac{N\sqrt{eB}}{\pi }\Big\{-\frac{M_{0}}{\sqrt{eB}}+\frac{\zeta
(1/2)}{\sqrt{2}}+\frac{\sqrt{eB}}{4T}+\frac{\sqrt{eB}}{2\sqrt{\pi }}%
\int_{0}^{\infty }\frac{dt}{t^{1/2}}\left( \coth {eBt}-1\right)  \nonumber \\
&\times &\left[ \Theta _{4}\Big(0|\frac{i}{4\pi T^{2}t}\Big)-1\right] \Big\},
\label{M_function} \\
\lambda (T,B) &=&\frac{N}{\pi \sqrt{eB}}\Big\{\frac{\zeta (3/2)}{4\sqrt{2}}+
\frac{1}{48}\left( \frac{\sqrt{eB}}{T}\right) ^{3}+\frac{(eB)^{3/2}}{2\pi }
\int_{0}^{\infty }dtt^{1/2}\left( \coth {eBt}-1\right)  \nonumber \\
&\times &\left[ \Theta _{4}\Big(0|\frac{i}{4\pi T^{2}t}\Big)-1\right] \Big\}.
\label{lambda_func}
\end{eqnarray}
The critical curve $M(T,B)=0$ can be analyzed analytically at $T\ll \sqrt{eB}
$ where the integral in Eq.(\ref{M_function}) is exponentially small and for
$T_{c}$ we get the equation
\begin{equation}
\frac{\sqrt{eB}}{4T_{c}}=\frac{M_{0}}{\sqrt{eB}}-\frac{\zeta (1/2)}{\sqrt{2}}%
.
\end{equation}
For the solution to exist, it must satisfy that $\sqrt{eB}\ll M_{0}$, what
gives the critical temperature
\begin{equation}
T_{c}\simeq \frac{eB}{4M_{0}}=\frac{1}{2}\sigma _{0},
\end{equation}
where $\sigma _{0}\equiv \sigma (T=0)$ is the dynamical mass at zero
temperature.

One can convince oneself that there is no solution of the equation $M(T,B)=0$
when $T_{c}\gg \sqrt{eB}$. Indeed, for that let us write the integral in (%
\ref{M_function}) as
\begin{eqnarray}
I &=&\int_{0}^{\infty }\frac{dt}{t^{1/2}}\left( \coth {eBt}-1\right) \left[
\Theta _{4}\Big(0|\frac{i}{4\pi T^{2}t}\Big)-1\right]  \nonumber \\
&=&\frac{1}{\sqrt{eB}}\int_{0}^{\infty }\frac{dt}{t^{1/2}}\left( \coth {t}
-1\right) \left[ \Theta _{4}\Big(0|\frac{ieB}{4\pi T^{2}t}\Big)-1\right] .
\label{I_integral}
\end{eqnarray}
We further divide the integrand into three pieces
\begin{eqnarray}
I &=&\frac{1}{\sqrt{eB}}\left\{ \int_{0}^{\infty }\frac{dt}{t^{3/2}}\left[
\Theta _{4}\Big(0|\frac{ieB}{4\pi T^{2}t}\Big)-1\right] -\int_{0}^{\infty }
\frac{dt}{t^{1/2}}\left( \coth {t}-\frac{1}{t}-1\right) \right.  \nonumber \\
&+&\left. \int_{0}^{\infty }\frac{dt}{t^{1/2}}\left( \coth {t}-\frac{1}{t}
-1\right) \Theta _{4}\Big(0|\frac{ieB}{4\pi T^{2}t}\Big)\right\} .
\end{eqnarray}
The first and second integrals in the last expression can be evaluated
exactly (after changing the variable $t\to x^{2}$ in the first integral)
with the help of the formulas \cite{Prudnikov}
\begin{eqnarray}
&&\int_{0}^{\infty }\frac{dt}{t^{3/2}}\left[ \Theta _{4}\Big(0|\frac{ieB}{
4\pi T^{2}t}\Big)-1\right] =-4\sqrt{\pi }\log {2}\frac{T}{\sqrt{eB}}; \\
&&\int_{0}^{\infty }\frac{dt}{t^{1/2}}\left( \coth {t}-\frac{1}{t}-1\right)
= \sqrt{2\pi }\zeta (\frac{1}{2}).
\end{eqnarray}
The integral with $\Theta _{4}-$function is calculated using the Jacobi
imaginary transformation to $\Theta _{2}-$function and keeping in it only
the first term in the series when $eB\to 0$:
\[
\Theta _{4}\Big(0|\frac{ieB}{4\pi T^{2}t}\Big)=\sqrt{\frac{4\pi T^{2}t}{eB}}
\Theta _{2}\Big(0|\frac{4i\pi T^{2}t}{eB}\Big)\simeq 4\sqrt{\frac{\pi T^{2}t
}{eB}}e^{-\frac{\pi ^{2}T^{2}t}{eB}}.
\]
This reduces the third integral to
\begin{eqnarray}
&&\int_{0}^{\infty }\frac{dt}{t^{1/2}}\left( \coth {t}-\frac{1}{t}-1\right)
\Theta _{4}\Big(0|\frac{ieB}{4\pi T^{2}t}\Big)\simeq \frac{4T\sqrt{\pi }}{%
\sqrt{eB}}\int_{0}^{\infty }{dt}\left( \coth {t}-\frac{1}{t}-1\right)
\nonumber \\
&&\times e^{-\frac{\pi ^{2}T^{2}t}{eB}}\simeq -\frac{4\sqrt{\pi eB}}{\pi
^{2}T},\quad B\to 0
\end{eqnarray}
(when proceeding to the last equality we changed the variable $t$ to $eBt$
and then expanded over $eB$).

Thus, combining all formulas we get the following expression for the $M-$
function:
\[
M(T,B)\simeq \frac{N\sqrt{eB}}{\pi }\left\{ -\frac{M_{0}}{\sqrt{eB}}+\frac{%
\sqrt{eB}}{4T}-2\log 2\frac{T}{\sqrt{eB}}-\frac{2}{\pi ^{2}}\frac{\sqrt{eB}}{
T}\right\} .
\]
As seen, there is no solution as $eB\to 0$ for $M_{0}>0$. (In case $eB=0$
and $M_{0}<0$ we get the standard expression for the critical temperature $%
T_{c}=|M_{0}|/{2\log 2}$ \cite{Rosenstein}). Hence, we arrive at the
conclusion that the only remaining possibility is that the root of the
equation $M(T,B)=0$ is of the order of $T_{c}\simeq \sqrt{eB}$. However, in
this case we cannot expand the integral in Eq.(\ref{M_function}) and should
turn to a numerical calculation. The root of the function $M(T,B)$ when the
parameter $M_{0}\simeq 0$ is found to be $\sqrt{eB}/T\simeq 4.1476$ what
defines the critical line. For the critical temperature this gives $%
T_{c}\simeq 0.54\sigma _{0}$ in agreement with the result of Ref.\cite{njl3}
. We calculated numerically the coefficient before $eB-eB_{c}$ in Eq. (\ref
{near_crit_scaling}) which is found to be $0.2738$, thus the scaling of the
dynamical mass near the critical line is given by the formula
\begin{equation}
\sigma \simeq 0.523\sqrt{eB-eB_{c}}.
\end{equation}

\section{Kubo Formula}

\label{App-B}

For the sake of completeness we derive here the expression for the thermal
conductivity in the two-dimensional case used in Section II. We start from
the Kubo's formula for the thermal conductivity tensor \cite{Langer}

\begin{equation}
\kappa _{ij}(\omega )=\frac{1}{VT}\int\limits_{0}^{\infty}dt
\int\limits_{0}^{^{\beta }}d\lambda {\rm Tr}\{\rho U_{j}(0)U_{i}(t+i\lambda
)\}e^{-i\omega t},  \label{1a}
\end{equation}
where $V$ is the space volume, $T$ the absolute temperature, $\rho$ is the
density matrix, and $U_{i}$ are the heat current operators with

\begin{equation}
U_{i}(t)=e^{iHt}U_{i}e^{-iHt}.  \label{3a}
\end{equation}
Integrating over $t$ by parts in Eq.(\ref{1a}), and taking into account that
the currents go to zero at $t\rightarrow \infty $, we obtain
\begin{eqnarray}
\kappa _{ij}(\omega )&=&\frac{1}{VT}\int\limits_{0}^{\infty }dt\frac{
e^{-i\omega t}-1}{i\omega }\int\limits_{0}^{^{\beta }}d\lambda \frac{
\partial }{\partial t}{\rm Tr}\{\rho U_{j}(0)U_{i}(t+i\lambda )\}  \nonumber
\\
&=&\frac{1}{VT}\int\limits_{0}^{\infty }dt\frac{e^{-i\omega t}-1}{\omega }%
{\rm Tr}\{\rho U_{j}(0)[U_{i}(t)-U_{i}(t+i\beta )]\},  \label{6a}
\end{eqnarray}
where we used also the fact that the quantity under ${\rm Tr}$ is a function
only of $t+i\lambda$.

Now, taking into account that

\begin{equation}
{\rm Tr}\{\rho U_{j}(0)U_{i}(t+i\beta )\}={\rm Tr}\{\frac{1}{Z}e^{-\beta
H}U_{j}(0)e^{iHt-\beta H}U_{i}(0)e^{-iHt+\beta H}\}={\rm Tr}\{\rho
U_{i}(t)U_{j}(0)\},  \label{7a}
\end{equation}
we obtain from (\ref{6a}) and (\ref{7a}) the known expression \cite{Verboven}

\begin{equation}
\kappa _{ij}(\omega )=-\frac{1}{VT}\int\limits_{0}^{\infty }dt\frac{%
e^{-i\omega t}-1}{\omega }{\rm Tr}\{\rho [U_{i}(t),U_{j}(0)]\}.  \label{8a}
\end{equation}
Using that

\begin{equation}
{\rm Tr}\{\rho U_{i}(t)_{j}(0)\}^{\dagger }={\rm Tr}\{U_{j}^{\dagger}(0)
U_{i}^{\dagger }(t)\rho^{\dagger }\}={\rm Tr}\{\rho U_{j}(0)U_{i}(t)\}
\label{10a}
\end{equation}
we have
\begin{eqnarray}
{\rm Tr}\{\rho[U_{i}(t),U_{j}(0)]\}={\rm Tr}\{\rho U_{i}(t)U_{j}(0)\}- {\rm %
Tr}\{\rho U_{i}(t)U_{j}(0)\}^{\dagger } =2i{\rm Im}{\rm Tr}\{\rho
U_{i}(t)U_{j}(0)\}.  \label{11a}
\end{eqnarray}
Then, we can write (\ref{8a}) as

\begin{equation}
\kappa _{ij}(\omega )=-\frac{2i}{VT}\int\limits_{0}^{\infty }dt\frac{%
e^{-i\omega t}-1}{\omega }{\rm Im}{\rm Tr}\{\rho U_{i}(t)U_{j}(0)\}.
\label{12a}
\end{equation}
The thermal conductivity for an isotropic system is given by $\kappa=\kappa
_{ii}(0)/d$ where the summation over repeated indices is understood ($d$ is
the number of space dimensions, in our case $d=2$). Equation (\ref{12a})
then takes the form

\begin{eqnarray}
\kappa =-\frac{i}{VT}\int\limits_{0}^{\infty}dt\lim_{\omega \rightarrow 0} (%
\frac{e^{-i\omega t}-1}{\omega}){\rm Im}{\rm Tr}\{\rho U_{i}(t)U_{i}(0)\} =-%
\frac{1}{VT}{\rm Im}\int\limits_{0}^{\infty}dtt {\rm Tr}\{\rho
U_{i}(t)U_{i}(0)\},  \label{14a}
\end{eqnarray}
which is equivalent to Eq.(\ref{heat_cur_cor}), or,

\begin{equation}
\kappa =\frac{i}{2VT}\int\limits_{0}^{\infty}dtt {\rm Tr}\{\rho
[U_{i}(t),U_{j}(0)]\}.  \label{14aa}
\end{equation}
In the representation of the Hamiltonian eigenfunctions, $e^{-iHt}\mid
n>=e^{-iE_{n}t}\mid n>$ we can write

\begin{eqnarray}
{\rm Tr}\{\rho U_{i}(t)U_{i}(0)\}&=&\sum_{n,m}\frac{1}{Z}\{e^{iHt+\beta
H}U_{i}e^{-iHt}\mid n><n\mid U_{i}\mid m><m\mid \}  \nonumber \\
&=&\sum_{n,m}\frac{1}{Z}e^{-\beta E_{n}+i(E_{n}-E_{m})t}\mid <n\mid
U_{i}\mid m>\mid ^{2},  \label{16a}
\end{eqnarray}
where the hermiticity of the heat current operators was used. Similarly
\begin{equation}
{\rm Tr}\{\rho U_{i}(0)U_{i}(t)\}=\sum_{n,m}\frac{1}{Z}e^{-\beta
E_{n}+i(E_{m}-E_{n})t}\mid <n\mid U_{i}\mid m>\mid ^{2}.  \label{18a}
\end{equation}
Now using Eqs.(\ref{16a}) and (\ref{18a}) and the symmetry of the matrix
elements under the interchange $n\leftrightarrow m$ we can write the
correlator function in the form

\begin{equation}
{\cal G}(t)={\rm Tr}\{\rho [U_{i}(t),U_{i}(0)]\}=\frac{1}{Z}%
\sum_{n,m}e^{-\beta E_{n}+i(E_{n}-E_{m})t}\left( 1-e^{-\beta
(E_{m}-E_{n})}\right) \mid <n\mid U_{i}\mid m>\mid ^{2}  \label{19a}
\end{equation}
The retarded Fourier transform of ${\cal G}(t)$ is given by
\begin{eqnarray}
{\cal G}(\Omega ) &=&\frac{1}{2\pi }\lim_{\eta \rightarrow
0}\int\limits_{-\infty }^{\infty }\theta \left( t\right) {\cal G}%
(t)e^{i\Omega t-\eta \left| t\right| }dt  \nonumber \\
&=&\frac{1}{2\pi }\frac{1}{Z}\sum_{n,m}e^{-\beta E_{n}}\left( 1-e^{-\beta
(E_{m}-E_{n})}\right) \mid <n\mid U_{i}\mid m>\mid ^{2}\lim_{\eta
\rightarrow 0}\int\limits_{0}^{\infty }dte^{i(E_{n}-E_{m}+\Omega )t-\eta t}
\nonumber \\
&=&\frac{1}{2\pi i}\frac{1}{Z}\lim_{\eta \rightarrow 0}\sum_{n,m}e^{-\beta
E_{n}}\left( 1-e^{-\beta (E_{m}-E_{n})}\right) \mid <n\mid U_{i}\mid m>\mid
^{2}\frac{1}{E_{m}-E_{n}-\Omega -i\eta }.  \label{21a}
\end{eqnarray}
To obtain the spectral representation of ${\cal G}(\Omega )$ we define
\begin{eqnarray}
\Theta (\Omega ) &=&\lim_{\eta \rightarrow 0}\left[ {\cal G}(\Omega +i\eta )-%
{\cal G}(\Omega -i\eta )\right]   \nonumber \\
&=&\sum_{n,m}\frac{1}{Z}e^{-\beta E_{n}}\left( 1-e^{-\beta
(E_{m}-E_{n})}\right) \mid <n\mid U_{i}\mid m>\mid ^{2}\delta
(E_{m}-E_{n}-\Omega ).  \label{22a}
\end{eqnarray}
From (\ref{21a}) and (\ref{22a}) we can write
\begin{equation}
{\cal G}(\Omega )=\lim_{\eta \rightarrow 0}\frac{1}{2\pi i}%
\int\limits_{-\infty }^{\infty }\Theta (\nu )\frac{d\nu }{\nu -\Omega -i\eta
}.  \label{23a}
\end{equation}
We can use this representation in order to express the thermal conductivity
in terms of the spectral density $\Theta (\Omega )$, which is an important
step for the calculations in the Green's function formalism. Using the
inverse Fourier transform, we can write

\begin{equation}
{\cal G}(t)=\int\limits_{-\infty }^{\infty }{\cal G}(\Omega )e^{-i\Omega
t}d\Omega =\lim_{\eta \rightarrow 0}\frac{1}{2\pi i}\int\limits_{-\infty
}^{\infty }e^{-i\Omega t}d\Omega \int\limits_{-\infty }^{\infty }\Theta (\nu
)\frac{d\nu }{\nu -\Omega -i\eta },\qquad t>0,  \label{24a}
\end{equation}
and considering the integration formula

\begin{equation}
\int\limits_{-\infty }^{\infty }\frac{e^{i\omega t}}{\omega -i\eta }d\omega
=\left\{
\begin{array}{ll}
2\pi ie^{-\eta t}, & t>0 \\
0, & t<0
\end{array}
\right. ,  \label{25a}
\end{equation}
the function ${\cal G}(t)$ can be transformed to

\begin{eqnarray}
{\cal G}(t)&=&\lim_{\eta \rightarrow 0}\frac{1}{2\pi i}\int\limits_{-\infty
}^{\infty }\Theta (\nu )d\nu \int\limits_{-\infty }^{\infty }\frac{
e^{-i\Omega t}}{\nu -\Omega -i\eta }d\Omega =\lim_{\eta \rightarrow 0}\frac{1%
}{2\pi i}\int\limits_{-\infty }^{\infty }\Theta (\nu )e^{-i\nu t}d\nu
\int\limits_{-\infty }^{\infty }\frac{ e^{i\omega t}}{\omega -i\eta }d\omega
\nonumber \\
&=&\lim_{\eta \rightarrow 0}\int\limits_{-\infty }^{\infty }\Theta (\Omega
)e^{-i\Omega t-\eta t}d\Omega ,\qquad t>0 .  \label{26a}
\end{eqnarray}
To express $\kappa $ in terms of the spectral function we substitute with (%
\ref{26a}) in (\ref{14aa}), so we get
\begin{eqnarray}
\kappa =\frac{i}{2VT}\int\limits_{0}^{\infty }dtt{\cal G} (t)=\frac{i}{2VT}
\int\limits_{0}^{\infty }dtt\lim_{\eta \rightarrow 0}\int\limits_{-\infty
}^{\infty }\Theta (\Omega )e^{-i\Omega t-\eta t}d\Omega =\frac{i}{2VT}
\lim_{\eta \rightarrow 0}\int\limits_{-\infty }^{\infty }\Theta (\Omega
)d\Omega \int\limits_{0}^{\infty }dtte^{-i\Omega t-\eta t}.  \label{27a}
\end{eqnarray}
Taking into account that the spectral function is an odd function, $\Theta
(\nu )=-\Theta (-\nu ),$ we can see that only the imaginary part of $%
\int\limits_{0}^{\infty }dte^{-i\Omega t-\eta t}t$ remains, what leads to
\begin{equation}
\kappa = -\frac{\pi }{2VT}\lim_{\eta \rightarrow 0}\int\limits_{-\infty
}^{\infty }\Theta (\Omega )\frac{\partial \delta (\Omega )}{\partial \Omega }
d\Omega =\frac{\pi }{2VT}\frac{\partial \Theta (\Omega )}{\partial \Omega }
\left.\Big|_{\Omega =0}\right.  \label{29a}
\end{equation}
or equivalently,

\begin{equation}
\kappa =\frac{\pi }{4VT}\lim_{\Omega \rightarrow 0}\frac{1}{\Omega }\left[
\Theta (\Omega )-\Theta (-\Omega )\right] .  \label{30a}
\end{equation}
From the representation (\ref{22a}) for the spectral function it can be
shown that the thermal conductivity (\ref{30a}) can be expressed in terms of
imaginary time Green's functions. Indeed, let us introduce the following
thermal Green function
\begin{equation}
\Pi \left( \tau \right) ={\rm Tr}\{\rho e^{H\tau }U_{i}(0)e^{-H\tau
}U_{i}(0)\}  \label{31a}
\end{equation}
and its Fourier transform
\begin{equation}
\Pi \left( i\omega _{n}\right) =\int\limits_{0}^{\beta }\Pi \left( \tau
\right) e^{i\omega _{n}\tau }d\tau ,\qquad \omega _{n}=\frac{2\pi n}{\beta }%
,\qquad n=0,1,2,...\,.  \label{33a}
\end{equation}
Inserting the complete set of energy eigenstates $|m\rangle ,|n\rangle $ we
can perform the integration over $\tau $ as indicated in (\ref{33a}), to
find
\begin{equation}
\Pi (i\omega _{k})=\frac{1}{Z}\sum_{n,m}e^{-\beta E_{n}}\mid <n\mid
U_{i}\mid m>\mid ^{2}\frac{e^{-(E_{m}-E_{n})\beta }-1}{E_{n}-E_{m}+i\omega
_{k}}.
\end{equation}
If we now define an analytical function $\Pi (\omega )$ in such a way that
at discrete points $\omega =i\omega _{k}$ it coincides with $\Pi (i\omega
_{k})$ and has a branch cut along the real axis, then the spectral function (%
\ref{22a}) is related to the discontinuity of $\Pi (\omega )$ across the cut
\begin{equation}
\Theta (\Omega )=\frac{1}{2\pi i}\lim_{\varepsilon \rightarrow 0}\left[ \Pi
\left( \Omega +i\varepsilon \right) -\Pi \left( \Omega -i\varepsilon \right)
\right] .  \label{32a}
\end{equation}
Substituting (\ref{32a}) in (\ref{30a}) we arrive to Eq. (10) of Section II.


\begin{references}
\bibitem{prl94}  V.~P.~Gusynin, V.~A.~Miransky, and I.~A.~Shovkovy, \prl
{\bf 73}, 3499 (1994). 

\bibitem{njl3}  V.~P.~Gusynin, V.~A.~Miransky, and I.~A.~Shovkovy, \prd
{\bf 52}, 4718 (1995). 

\bibitem{njl4}  V.~P.~Gusynin, V.~A.~Miransky, and I.~A.~Shovkovy, \pl B
{\bf 349}, 477 (1995). 

\bibitem{Klimenko}  K.~G.~Klimenko, Z. Phys. C{\bf 54}, 323 (1992);
Teor. Mat. Fiz. {\bf 90}, 3 (1992); 
I.~V.~ Krive and S.~A.~Naftulin, Phys. Rev. D {\bf 46}, 2737 (1992).

\bibitem{qed4}  V.~P.~Gusynin, V.~A.~Miransky, and I.~A.~Shovkovy, \prd
{\bf 52}, 4747 (1995); 
Nucl. Phys. B {\bf 462}, 249 (1996); 
\prl {\bf 83}, 1291 (1999); 
Nucl. Phys. B {\bf 563}, 361 (1999); 
V.~Elias {\sl et al.}, Phys. Rev. D {\bf 54}, 7884 (1996);
V.~ P.~Gusynin and A.~V.~Smilga, Phys.Lett. B {\bf 450}, 267 (1999).

\bibitem{misc}  C. N. Leung, Y. J. Ng, and A. W. Ackley, \prd {\bf 54}, 4181
(1996); 
D. K. Hong, Y. Kim, and S.-J. Sin, \prd {\bf 54}, 7879 (1996);
A. V. Shpagin, hep-ph/9611412; 
D.~S.~Lee , C.~N.~Leung, and Y.~J.~Ng, Phys. Rev. D {\bf 55}, 6504 (1997);
V.~P.~Gusynin and I.~A.~Shovkovy, Phys. Rev. D {\bf 56}, 5251 (1997).

\bibitem{other}  I.~A.~Shushpanov and A.~V.~Smilga, \pl B {\bf 402}, 351
(1997); 
S.~Kanemura, H.~-T.~Sato and H.~Tochimura, Nucl. Phys. B {\bf 517}, 567
(1998); 
A.~Yu.~Babansky, E.~V.~Gorbar, and G.~V.~Shchepanyuk, \pl B {\bf 419}, 272
(1998); 
T.~Itoh and H.~Kato, \prl {\bf 81}, 30 (1998); E.~J.~Ferrer and V. de la
Incera, Phys. Lett. B {\bf 481},287 (2000); 
V.~P.~Gusynin, Ukrainian J. Phys. {\bf 45}, 603 (2000);
V.~Ch.~Zhukovsky {\sl et al.}, JETP Lett. {\bf 73}, 121 (2001).

\bibitem{Nick}  K. Farakos and N.E. Mavromatos, cond-mat/9710288.

\bibitem{Nick2}  K.~Farakos, G.~Koutsoumbas, and N.E.~Mavromatos, Int. J.
Mod. Phys. B{\bf 12}, 2475 (1998). 

\bibitem{Semenoff}  G.~W.~Semenoff, I.~A.~Shovkovy, and
L.~C.~R.~Wijewardhana, Mod. Phys. Lett. A{\bf 13}, 1143 (1998).

\bibitem{Liu}  W.~V.~Liu, Nucl. Phys. B {\bf 556}, 563 (1999).

\bibitem{Ferrer}  E.~J.~Ferrer, V.~P.~Gusynin, and V. de la Incera, Mod.
Phys. Lett. B{\bf 16}, 107 (2002). 

\bibitem{applic}  E.~J.~Ferrer, V.~P.~Gusynin, and V. de la Incera, Phys.
Lett. B {\bf 455}, 217 (1999); 
G.~W.~Semenoff, I.~A.~Shovkovy, and L.~C.~R.~Wijewardhana, Phys. Rev. D {\bf %
\ 60}, 105024 (1999). 

\bibitem{Krishana}  K.~Krishana {\sl et al.}, Science, {\bf 277}, 83 (1997),
N.~P.~Ong {et al.}, cond-mat/9904160.

\bibitem{Aubin}  H.~Aubin {\sl et al.}, Phys. Rev. Lett., {\bf 82}, 624
(1999). 

\bibitem{Ando}  Y.~Ando et al., Phys. Rev. B {\bf 62}, 626 (2000).

\bibitem{Ando2}  Y.~Ando et al., Phys. Rev. Lett. {\bf 88}, 147004 (2002).

\bibitem{K-2}  D.~V.~Khveshchenko, Phys. Rev. Lett. {\bf 87}, 206401 (2001).

\bibitem{graphite}  M.~S.~Sercheli et al., cond-mat/0106232.

\bibitem{Rosenstein}  B.~Rosenstein and B.~Warr, Phys. Repts. {\bf 205}, 59
(1991). 

\bibitem{Ambegaokar}  V.~Ambegaokar and A.~Griffin, Phys. Rev. A {\bf 137},
1151 (1965). 

\bibitem{Lee}  A.~C.~Durst and P.~A.~Lee, Phys. Rev. B {\bf 62}, 1270
(2000). 

\bibitem{Chodos}  A.~Chodos, K.~Everding, and D.~A.~Owen, Phys. Rev. D {\bf %
42}, 2881 (1990). 

\bibitem{Franz}  M.~Franz, Phys. Rev. Lett. {\bf 82}, 1760 (1999).

\bibitem{univ-kappa_exp}  L.~Taillefer et al., Phys. Rev. Lett. {\bf 79},
483 (1997); 
K.~Behnia et al., J. Low Temp. Phys. {\bf 117}, 1089 (1999).

\bibitem{Franz2}  M.~Franz and O.~Vafek, \prb {\bf 64}, 220501(R) (2001).

\bibitem{Anderson}  P.~W.~Anderson, cond-mat/9812063.

\bibitem{Vekhter}  I.~Vekhter and A.~Houghton, Phys. Rev. Lett. {\bf 83},
4626 (1999). 

\bibitem{Gonzales}  G. Semenoff, Phys. Rev. Lett. {\bf 53}, 2449 (1984);
F.~D.~M.~ Haldane, {\sl ibid} {\bf 61}, 2015 (1988); J.~Gonzales, F.~Guinea,
and M.~A.~H.~Vozmediano, Phys. Rev. Lett. {\bf 77}, 3589 (1996).

\bibitem{Volovik}  G.~E.~Volovik, Proc. Nat. Acad. Sci. {\bf 96}, 6042
(1999); 
Phys. Rept. {\bf 351} 195 (2001). 

\bibitem{GGMS}  E.V.~Gorbar, V.P.~Gusynin, V.A.~Miransky, and I.A.~Shovkovy,
cond-mat/0202422. To be published in PRB.

\bibitem{condmatdirac}  G.~W.~Semenoff and L.~C.~R.~Wijewardhana, \prl
{\bf 63}, 2633 (1989); 
J.~B.~Marston, Phys. Rev. Lett. {\bf 64}, 1166 (1990);
N.~Dorey and N.~E.~Mavromatos, Nucl. Phys. B {\bf 368}, 614 (1992);
S.~H.~Simon and P.~A.~Lee, \prl {\bf 78}, 1548 (1997). 

\bibitem{Randeria}  A.~Paramekanti {\sl et al.}, \prb {\bf 62}, 6786 (2000).

\bibitem{Laughlin}  R.~B.~Laughlin, \prl  {\bf 80}, 5188 (1998);
T.~V.~Ramakrishnan, J. Phys. Chem. Solids, {\bf 59}, 1750 (1998).

\bibitem{Melnikov}  A.~S.~Mel'nikov J. Phys. Cond. Matter {\bf 11}, 4219
(1999).

\bibitem{Tesa}  M.~Franz and Z.~Tesanovic, Phys. Rev. Lett. {\bf 84}, 554
(2000).

\bibitem{F-T-approach}  L.~Marinelli et al., Phys. Rev. B {\bf 62}, 3499
(2000); 
J.~Ye, Phys. Rev. Lett., {\bf 86}, 316 (2001); 
O.~Vafek et al, Phys. Rev. B {\bf 63}, 134509 (2001); 
J.~Ye and A.~Millis, cond-mat/0101032; 
D.~Knapp, C.~Kallin and A.~J.~Berlinsky, Phys. Rev. B, 64, 014502 (2001);
A. ~Vishwanath, Phys. Rev. Lett. 87, 217004 (2001).

\bibitem{Chakra}  S.~ Chakravarty, R.~ B.~ Laughlin, D.~ K.~ Morr, and C.~
Nayak,Phys. Rev. B {\bf 63}, 094503 (2001).

\bibitem{Nayak}  X.~ Yang and C.~ Nayak, Phys. Rev. B {\bf 65}, 064523
(2001).

\bibitem{Prudnikov}  A.~P.~Prudnikov, Yu.~A.~Brychkov and O.~I.~Marichev,
Integrals and Series, v.III, Nauka, Moscow, 1986. 

\bibitem{Langer}  J.~S.~Langer, Phys. Rev. {\bf 127}, 5 (1962).

\bibitem{Verboven}  E.~Verboven, Physica {\bf 26}, 1091 (1960).
\end{references}
\end{document}